\documentclass[reprint,superscriptaddress,nofootinbib,amsmath,amssymb,aps,prd]{revtex4-2}

\usepackage[T1]{fontenc}
\usepackage[colorlinks=true, allcolors=Blue]{hyperref}
\usepackage{microtype}
\usepackage{graphicx}
\usepackage[dvipsnames]{xcolor}
\usepackage{tensor}
\usepackage[capitalise]{cleveref}

\crefname{section}{Appendix}{Appendices}
\Crefname{section}{Appendix}{Appendices}
\crefname{subsection}{Appendix}{Appendices}
\Crefname{subsection}{Appendix}{Appendices}

\def\dd{\mathrm{d}}

\newcommand{\sect}[1]{\emph{#1}---}

\begin{document}

\title{Analytic solutions for the gravitational spin Hall equations and black hole shadows}

\author{Marius A. Oancea}
\email{marius.oancea@univie.ac.at}
\affiliation{University of Vienna, Faculty of Physics, Boltzmanngasse~5, 1090 Vienna, Austria}

\begin{abstract}
In general relativity, the geometric optics approximation describes the propagation of high-frequency electromagnetic waves along null geodesics. Beyond this approximation, localized wave packets acquire frequency- and polarization-dependent corrections to null geodesic motion, described by the gravitational spin Hall equations. In this paper, these equations are integrated analytically, to first order in wavelength, in a large class of type D black hole spacetimes, whenever the corresponding null geodesics admit analytic solutions. The same framework yields analytic expressions for the shape of black hole shadows with polarization-dependent corrections.
\end{abstract}

\maketitle

\sect{Introduction}Spin Hall effects arise from spin–orbit coupling and produce spin-dependent transverse transport in condensed matter physics \cite{SHE_review,PhysRevLett.92.126603,originalSHE1,originalSHE2,originalSHE3,originalSHE4} and optics \cite{SOI_review,SHEL_review,SHE-L_original,SHE_original,Bliokh2009,Duval2006,Duval2007,Bliokh2008,SHEL_experiment,OpticalMagnus,SHE_math}. In the optical case, electromagnetic wave packets undergo polarization-dependent deflections in inhomogeneous media or shifts upon reflection and refraction at sharp interfaces. These effects have been measured in several experiments \cite{SHEL_experiment,Bliokh2008,Qin:09,PradeepChakravarthy:19,LOFFLER20103367}.

The gravitational spin Hall effect \cite{GSHE_rev} is the general-relativistic analog of this phenomenon, with spacetime curvature playing the role of the optical inhomogeneity. Beyond geometric optics, wave-packet trajectories acquire frequency- and spin-dependent corrections to geodesic motion. This effect has been derived for electromagnetic waves \cite{GSHE2020,HarteOancea,SHE_QM1,Frolov2020,PhysRevD.110.064020,Frolov_2024a}, gravitational waves \cite{GSHE_GW,SHE_GW,Frolov_2024}, and Dirac fields \cite{audretsch,rudiger,GSHE_Dirac,SHE_Dirac}.

Polarization-dependent gravitational lensing offers new opportunities to probe strong-field gravity and the environments of compact sources. For lensed gravitational waves from hierarchical triple black hole systems, spin Hall corrections can leave potentially observable imprints on arrival times and waveforms \cite{GSHE_lensing,GSHE_lensing2}. Complementary spin-optics and wave-optics studies have also revealed polarization-dependent signatures in gravitational-wave lensing \cite{PhysRevD.109.044027}. Recent wave-optical calculations further exhibit helicity-dependent displacements in images of electromagnetic sources near black holes \cite{Nazare2026}, providing a benchmark for polarization-dependent propagation beyond geometric optics.

Using the conservation laws derived in \cite{GSHE_int}, this paper derives analytical solutions for the gravitational spin Hall equations in a large class of type D spacetimes. A strategy analogous to that used for massive spinning particles \cite{SkoupyWitzany2025} reduces the spin Hall dynamics to auxiliary null geodesics, allowing known analytic geodesic solutions to be used directly. Polarization-dependent corrections to the shape of black hole shadows are also derived analytically by extending methods used in the geodesic case \cite{Grenzebach2014,BHS3,Perlick2021}.

\sect{Gravitational spin Hall equations}Consider a localized, circularly polarized electromagnetic wave packet propagating on a fixed spacetime background. A high-frequency approximation of Maxwell's equations, retaining the first-order corrections in wavelength, yields an effective description of its propagation beyond geometric optics \cite{GSHE2020}. The resulting dynamics is expressed in terms of the energy centroid and the average momentum of the wave packet and is governed by the gravitational spin Hall equations \cite{HarteOancea}:
\begin{subequations} \label{eq:spin_hall}
\begin{align}
    \dot x^\mu &= p^\mu + \frac{S^{\mu\nu}}{p\cdot t}\,p^\alpha \nabla_\alpha t_\nu,\\
    \frac{D p_\mu}{\dd\lambda} &= -\frac12 R_{\mu\nu\alpha\beta} p^\nu S^{\alpha\beta}. 
\end{align}
In these equations, the external degrees of freedom are represented by $x^\mu$, the energy centroid of the wave packet with respect to the timelike vector field $t$, and the average momentum of the wave packet $p_\mu$. The internal polarization degree of freedom of the wave packet is encoded in the spin tensor $S^{\mu \nu}$, defined as
\begin{equation}
    S^{\mu\nu} = \frac{\varepsilon s}{p\cdot t}\,\epsilon^{\mu\nu\alpha\beta} p_\alpha t_\beta,
    \label{eq:spin_tensor}
\end{equation}
\end{subequations}
where $s = \pm 1$ for circularly polarized electromagnetic wave packets, and $\varepsilon$ is a small expansion parameter related to the wavelength. In particular, if we denote by $\omega$ the wave frequency measured by a timelike observer with $4$-velocity $t^\alpha$, then we have $p \cdot t = - \varepsilon \omega$.  These equations are understood through first order in $\varepsilon$, with errors of order $\varepsilon^2$. The condition $0 \leq \varepsilon \ll 1$ refers to $\varepsilon$ measured in units of an appropriate length scale over which the background geometry varies significantly.

It has been shown in \cite{HarteOancea} that the gravitational spin Hall equations are a particular massless case of the Mathisson--Papapetrou--Dixon equations \cite{Dixon74,HarteGrav}. \Cref{eq:spin_hall} follows from the pole-dipole Mathisson--Papapetrou--Dixon equations by imposing the spin supplementary condition $S^{\mu \nu} t_\mu = 0$, together with the restrictions that $p_\mu p^\mu = 0$ and $S^{\mu \nu} p_\mu = 0$ hold, at least initially. Based on this, it has been shown that the spin Hall equations admit conserved quantities that generalize the known ones from the geodesic case \cite{GSHE_int}. These are
\begin{subequations}
\begin{align}
    H &= \frac{1}{2} g^{\mu \nu} p_\mu p_\nu = 0, \\
    \mathcal{C}_{\kappa} &= \kappa^\mu p_\mu + \frac{1}{2} S^{\mu \nu} \nabla_\mu \kappa_\nu, \\
    \mathcal{D} &= \mathcal{K}^{\mu \nu} p_\mu p_\nu + \mathcal{L}_{\alpha \beta \mu} S^{\alpha \beta} p^\mu, \label{eq:D_CKY}
\end{align}
\end{subequations}
where $\kappa^\mu$ is a conformal Killing vector, $\mathcal{K}^{\mu \nu} = h^{\mu \rho} h\indices{_\rho^\nu}$ is a conformal Killing tensor, $h_{\alpha \beta}$ is a conformal Killing-Yano tensor, and $\mathcal{L}_{\alpha\beta\mu} = 2 \left( h_{\rho[\alpha} \nabla^\rho h_{\beta]\mu} - h\indices{_\mu^\rho} \nabla_{[\alpha}h_{\beta\rho]} \right)$.
The conservation of these quantities should be understood in an approximate sense, up to error terms of order $\varepsilon^2$. These conservation laws have been used to show that the spin Hall equations \eqref{eq:spin_hall} are completely integrable in a large class of type D spacetimes \cite{GSHE_int}.

\sect{Carter spacetimes}We use geometrized units and consider the Carter metric in canonical coordinates,
\begin{align}
    g &= -\frac{\Delta_r}{\Sigma}(\dd\tau + y^2 \dd\psi)^2 + \frac{\Sigma}{\Delta_r} \dd r^2 + \frac{\Sigma}{\Delta_y} \dd y^2 \nonumber\\
    &\quad + \frac{\Delta_y}{\Sigma}(\dd\tau - r^2 \dd\psi)^2,
\end{align}
where $\Sigma = r^2 + y^2$, and $\Delta_r = \Delta_r(r)$ and $\Delta_y = \Delta_y(y)$ are arbitrary smooth functions. No gravitational field equations are imposed at this stage. We work in a regular coordinate region where $\Sigma>0$, $\Delta_r>0$, and $\Delta_y>0$, and choose the orthonormal coframe
\begin{align}
    e^0 &= \sqrt{\frac{\Delta_r}{\Sigma}}(\dd\tau + y^2 \dd\psi), & e^1 &= \sqrt{\frac{\Sigma}{\Delta_r}} \, \dd r, \nonumber\\
    e^2 &= \sqrt{\frac{\Sigma}{\Delta_y}} \, \dd y, & e^3 &= \sqrt{\frac{\Delta_y}{\Sigma}}(\dd\tau - r^2 \dd\psi),
\end{align}
and define the dual frame by $(e^a)_\mu (e_b)^\mu = \delta^a_b$. We denote frame components as $p^a=(e^a)_\mu p^\mu = (p^0, p^1, p^2, p^3)$ and coordinate components as $p^\mu = (p^\tau,p^r,p^y,p^\psi)$.

Carter spacetimes admit two Lie-commuting Killing vectors and a closed conformal Killing--Yano tensor \cite{Frolov2017}:
\begin{equation}
    k = \partial_\tau,\qquad m = \partial_\psi,\qquad h = r e^0 \wedge e^1 + y e^2 \wedge e^3.
\end{equation}
The Hodge dual of $h$ is the Killing--Yano tensor $f = \star h = y e^0 \wedge e^1 - r e^2 \wedge e^3$.

\sect{Analytic solutions for the spin Hall equations}We now consider the spin Hall equations \eqref{eq:spin_hall} in Carter spacetimes, and we show that a suitable change of observer reduces these equations to null geodesic equations, which can then be integrated by quadratures. To start, we use the Carter observer $t$ to define the wave-packet centroid:
\begin{equation}
    t = e_0 = \frac{r^2 \partial_\tau + \partial_\psi}{\sqrt{\Sigma\Delta_r}}.
\end{equation}
This observer respects both Killing symmetries, although this is not required for the four conservation laws $(H = 0, E = -\mathcal{C}_k, L = \mathcal{C}_m, \mathcal{D})$.

Our general strategy now is to introduce auxiliary observers to reduce the spin Hall equations to null geodesics, integrate these reduced equations, and then transform back to the Carter observer. This is similar to the analytic treatment for massive spinning particles in \cite{SkoupyWitzany2025}.

A change of observer from $t$ to $\tilde{t}$, or equivalently a change of the spin supplementary condition, shifts the description of the same wave packet from $(x, p, S)$ to $
(\tilde{x}, \tilde{p}, \tilde{S})$. Through first order in $\varepsilon$, the result is \cite{HarteOancea,Vines2016}
\begin{subequations}
\label{eq:observer_centroid_map}
\begin{align}
    \tilde{x} &= \exp_x(\delta x), \qquad \tilde{p}^{\mu'} = G^{\mu'}{}_{\nu} p^\nu, \\
    \tilde{S}^{\mu'\nu'} &= G^{\mu'}{}_{\alpha} G^{\nu'}{}_{\beta} \left(S^{\alpha\beta} + 2p^{[\alpha}\delta x^{\beta]}\right),
\end{align}
\end{subequations}
where $\exp$ is the exponential map, the bitensor $G$ is the parallel propagator to the displaced point, denoted by primed indices \cite{Poisson2011}, and the shift vector $\delta x$ is defined as
\begin{equation}
    \delta x^\mu = \frac{S^{\mu\nu}\tilde{t}_\nu}{p \cdot \tilde{t}}.
    \label{eq:observer_shift_vector}
\end{equation}
The conserved quantities $(H, E, L, \mathcal{D})$ are unchanged, up to error terms of order $\varepsilon^2$, as shown in \cref{app:observer_invariance}.

Differentiation of the transformations in \cref{eq:observer_centroid_map} gives, through first order, \cite{Vines2016}
\begin{subequations}

\begin{align}
    \frac{\dd \tilde{x}^{\mu'}}{\dd\lambda} &= G^{\mu'}{}_{\mu} \left(\dot x^\mu + \frac{D\delta x^\mu}{\dd\lambda}\right), \label{eq:observer_shifted_velocity}\\
    \frac{D\tilde{p}^{\mu'}}{\dd\lambda} &= G^{\mu'}{}_{\mu} \left(\frac{Dp^\mu}{\dd\lambda} + R^\mu{}_{\nu\alpha\beta} p^\nu \delta x^\alpha p^\beta\right). \label{eq:observer_shifted_force}
\end{align}
\end{subequations}
Thus, we seek an auxiliary observer $\tilde{t}$, such that the anomalous velocity and curvature terms in the above equations are removed or reduced to a simpler form.

We first consider the case where the leading-order part of $p_\mu$ is not aligned with a principal null direction, and thus $\mathcal{D} = \mathcal{O}(1)$. In this case, we consider an auxiliary observer 
\begin{align}
    \tilde{t}^\mu = \frac{h^\mu{}_{\rho} h^\rho{}_{\nu} + y^2 \delta^\mu_{\nu}}{\Sigma} p^\nu = p^0 (e_0)^\mu + p^1 (e_1)^\mu.
    \label{eq:adapted_observer}
\end{align}
Geometrically, $\tilde{t}$ is the projection of $p$ onto the Lorentzian principal plane spanned by $e_0$ and $e_1$, which contains both principal null directions. In this case, the shift vector takes the form
\begin{equation}
    \delta x^\mu = \frac{\varepsilon s\,p^1}{\mathcal D p^0} (r f^\mu{}_{\nu} - y h^\mu{}_{\nu}) p^\nu.
    \label{eq:carter_displacement}
\end{equation}
Note that $r f - y h = -\Sigma e^2 \wedge e^3$, and we have $p \cdot \delta x = t \cdot \delta x = \delta x^\mu\nabla_\mu r = 0$. Thus, the radial component is not affected by the shift. The magnitude of the shift vector
\begin{equation}
    \sqrt{g(\delta x,\delta x)} = \frac{|\varepsilon s p^1|}{|p^0| \sqrt{(p^0)^2 - (p^1)^2}} + \mathcal O(\varepsilon^2). 
\end{equation}
is finite for nonprincipal null directions, but is not bounded as a principal direction is approached at a fixed local frequency. The perturbative transformation is used where this displacement and the associated momentum correction remain small. The principal null case will be treated separately below.

To simplify the notation, primed spacetime indices are suppressed. All background quantities in the following equations are evaluated at the shifted centroid, and the displacement is parallel transported there when contracted with them. The shifted spin tensor is
\begin{equation}
    \tilde{S}^{\mu\nu} = -\frac{2\varepsilon s}{\mathcal D}  h^{[\mu}{}_{\alpha}f^{\nu]}{}_{\beta} \tilde{p}^\alpha \tilde{p}^\beta.
    \label{eq:adapted_spin}
\end{equation}
The shifted equations take the compact form
\begin{subequations}
\label{eq:adapted_gshe}
\begin{align}
    \frac{\dd \tilde{x}^\mu}{\dd\lambda} &= \Sigma N \hat{p}^\mu, \label{eq:adapted_velocity}\\
    \frac{\tilde{D}\tilde{p}^\mu}{\dd\lambda} &= -\frac{\varepsilon s}{\mathcal D} \left[\tilde{p}^\alpha \tilde{p}^\beta \nabla_\alpha \nabla_\beta(ry)\right] \tilde{p}^\mu, \label{eq:adapted_force}
\end{align}
\end{subequations}
where
\begin{subequations}
\begin{align}
    N &=  \frac{\Sigma+\delta x^\mu \nabla_\mu \Sigma}{\Sigma^2}  \left[1 - \frac{\varepsilon s \tilde{p}^\alpha(y \nabla_\alpha r - r \nabla_\alpha y)}{\Sigma(\tilde{p} \cdot t)^2} \right], \label{eq:carter_parameter_factor}\\
    \hat{p}^\mu &= \tilde{p}^\mu + \frac{\varepsilon s}{\mathcal D} \left[E f^\mu{}_{\nu} \tilde{p}^\nu + \tilde{p}^\alpha \nabla_\alpha(ry) \, \tilde{p}^\mu\right]. \label{eq:carter_auxiliary_momentum}
\end{align}
\end{subequations}
The derivation of these equations is given in \cref{app:shift_dynamics}.

Finally, we pass from the above equations to the reduced geodesic variables $(\hat{x} = \tilde{x}, \hat{p})$. The derivative of the momentum correction in \cref{eq:carter_auxiliary_momentum} cancels the remaining force term in \cref{eq:adapted_force}, as shown in \cref{app:shift_dynamics}, and the values of the conserved quantities are also preserved, $(\hat{H},\hat{E},\hat{L},\hat{\mathcal D}) = (H,E,L,\mathcal D) + \mathcal O(\varepsilon^2)$, as shown in \cref{app:observer_invariance}. We also introduce the generalized Mino parameter $\gamma$ defined by $\dd\gamma = N \dd\lambda$, which then leads to the reduced null geodesic equation
\begin{equation}
    \frac{\dd\hat{x}^\mu}{\dd\gamma} = \Sigma(\hat{x})\hat{p}^\mu, \qquad \frac{\hat{D}\hat{p}_\mu}{\dd\gamma} = 0.
    \label{eq:carter_auxiliary_geodesic}
\end{equation}
Introducing $P(\hat{r}) = \hat{r}^2 \hat{E} - \hat{L}$ and $Q(\hat{y}) = \hat{y}^2 \hat{E} + \hat{L}$, the geodesic equations separate:
\begin{subequations}
\label{eq:carter_separated}
\begin{align}
    \left(\frac{\dd \hat{r}}{\dd\gamma}\right)^2 &= P(\hat{r})^2 + \Delta_r(\hat{r})\hat{\mathcal D},\\
    \left(\frac{\dd\hat{y}}{\dd\gamma}\right)^2 &= -Q(\hat{y})^2 - \Delta_y(\hat{y})\hat{\mathcal D},\\
    \frac{\dd\hat{\psi}}{\dd\gamma} &= \frac{P(\hat{r})}{\Delta_r(\hat{r})} + \frac{Q(\hat{y})}{\Delta_y(\hat{y})},\\
    \frac{\dd\hat{\tau}}{\dd\gamma} &= \frac{\hat{r}^2 P(\hat{r})}{\Delta_r(\hat{r})} - \frac{\hat{y}^2Q(\hat{y})}{\Delta_y(\hat{y})}.
\end{align}
\end{subequations}
For arbitrary smooth Carter metric functions, \cref{eq:carter_separated} determines the auxiliary null trajectories by quadratures. If both metric functions are polynomials of degree at most four, the separated potentials are at most quartic, and the solutions can be expressed using elliptic functions and integrals \cite{10.1063/1.4861945}. Repeated roots or potentials of degree at most two yield elementary expressions for the corresponding quadratures. This polynomial condition does not require the field equations and includes the Kerr--Newman--NUT--(anti-)de Sitter family. Explicit solutions are given for Kerr \cite{GrallaLupsasca2020}, Kerr--(anti-)de Sitter \cite{Hackmann2010}, and the nonaccelerating Pleba\'nski--Demia\'nski family \cite{Hackmann2009}. Higher-degree polynomial potentials generally lead to hyperelliptic integrals \cite{ENOLSKI2011899}. Arbitrary smooth metric functions need not admit an elliptic or hyperelliptic representation.

Given null geodesic solutions to \cref{eq:carter_separated}, the spin Hall rays for the original Carter observer are obtained algebraically at the same value of the Mino parameter:
\begin{subequations}
\label{eq:carter_inverse_coordinates}
\begin{align}
    r(\gamma) &= \hat{r}(\gamma),\\
    y(\gamma) &= \hat{y}(\gamma) - \frac{\varepsilon s \, \Delta_r \hat{p}_r Q}{\hat{\mathcal D} P},\\
    \psi(\gamma) &= \hat{\psi}(\gamma) + \frac{\varepsilon s \, \Delta_r \hat{p}_r \hat{p}_y}{\hat{\mathcal D} P},\\
    \tau(\gamma) &= \hat{\tau}(\gamma) - \frac{\varepsilon s \, \Delta_r \hat{p}_r \hat{y}^2 \hat{p}_y}{\hat{\mathcal D} P}.
\end{align}
\end{subequations}
Every first-order correction is evaluated on the geodesic solution at that same parameter value. The required momentum components are already known from the geodesic solution:
\begin{equation}
    \hat{p}_r = \frac{1}{\Delta_r(\hat{r})} \frac{\dd \hat{r}}{\dd\gamma}, \qquad \hat{p}_y = \frac{1}{\Delta_y(\hat{y})} \frac{\dd \hat{y}}{\dd\gamma}.
    \label{eq:carter_auxiliary_covector_gamma}
\end{equation}
Together with the physical momentum components in \cref{eq:app_inverse_covector_components}, these formulas give the complete solution to the spin Hall equations through first order. Physical initial data determine the auxiliary initial data by the forward transformation. Once the geodesic is known, reconstruction involves only algebraic operations on its position and momentum, with no additional integration. Recovering the original parameter $\lambda$ is optional and requires the quadrature in \cref{eq:app_reconstruct_parameter} and its inversion.

For initial momenta aligned with the principal null directions, the auxiliary observer in \cref{eq:adapted_observer} becomes null. We instead use a timelike observer parallel propagated along the leading principal null geodesic. This transport can also be solved analytically \cite{Kubiznak2009,Marck1983Null} in Carter spacetimes, and the reduction and momentum reconstruction are given in \cref{app:principal_observer}. The auxiliary geodesic then has $\hat{y} = y_0$, $\Sigma_0 = r_0^2 + y_0^2$, where the subscript zero denotes initial data, and $\chi = \pm 1$ specifies the initial radial orientation. With the auxiliary affine momentum chosen to equal the physical momentum initially, the corresponding Carter trajectory is
\begin{subequations}
\label{eq:principal_inverse_coordinates}
\begin{align}
    r(\gamma) &= \hat{r}(\gamma),\\
    y(\gamma) &= y_0 - \frac{\varepsilon s \chi \, \Delta_y(y_0) r_0(\hat{r} - r_0)}{\hat{E} \Sigma_0 \Sigma(\hat{x})},\\
    \psi(\gamma) &= \hat{\psi}(\gamma) - \frac{\varepsilon s y_0(\hat{r} - r_0)}{\hat{E} \Sigma_0 \Sigma(\hat{x})},\\
    \tau(\gamma) &= \hat{\tau}(\gamma) + \frac{\varepsilon s y_0^3(\hat{r} - r_0)}{\hat{E} \Sigma_0 \Sigma(\hat{x})}.
\end{align}
\end{subequations}
All corrections are evaluated on the auxiliary principal geodesic. The physical and auxiliary constants obey the relations in \cref{eq:app_principal_cons}, and the physical momentum follows algebraically from \cref{eq:app_principal_inverse_momentum}. Initial momenta with a principal leading part and transverse components of order $\varepsilon$ are covered by the extension in \cref{app:principal_observer}, using an auxiliary geodesic with the full initial momentum.

These results can also be extended to conformally Carter metrics, as described in \cref{app:conformal_carter}.

\sect{Trapping and black hole shadows}We now turn to polarization-dependent effects on trapping and black hole shadows described by the spin Hall equations \eqref{eq:spin_hall} in Carter spacetimes. Using the conservation laws and extending the null-geodesic methods developed in \cite{Grenzebach2014,BHS3,Perlick2021}, we derive analytic expressions for the polarization-corrected shadow boundary.

Since $r(\gamma) = \hat r(\gamma)$, radial trapping is governed by the same potential and stability conditions as for the auxiliary null geodesics. In particular,
\begin{equation}
    \left(\frac{\dd\hat r}{\dd\gamma}\right)^2 = (\hat r^2\hat E-\hat L)^2+\Delta_r(\hat r)\hat{\mathcal D} = \mathcal R(\hat r),
\end{equation}
where $\mathcal R(r)=(r^2E-L)^2+\Delta_r(r)\mathcal D$ on the nonprincipal branch. Unstable spherical spin Hall rays have radii $r_{\rm p}$ satisfying
\begin{equation}
    \mathcal R(r_{\rm p}) = \mathcal R'(r_{\rm p}) = 0, \qquad \mathcal R''(r_{\rm p}) > 0.
    \label{eq:shadow_spherical_conditions}
\end{equation}
We assume a regular black hole exterior whose capture and escape regions are separated by accessible unstable spherical rays. These are restrictions on the off-shell metric functions, as in the usual shadow construction for nonaccelerating Pleba\'nski--Demia\'nski black holes \cite{Grenzebach2014,BHS7,Perlick2021}.

Following \cite[Sec.~IV]{Grenzebach2014}, consider a Carter observer at $(r_O,y_O)$ and write the future-directed momentum of a received photon in coordinate and frame components:
\begin{align}
    &p =  p^\tau\partial_\tau+p^r\partial_r+p^y\partial_y+p^\psi\partial_\psi  \nonumber\\
    &= p^0\big(e_0+\cos\alpha\,e_1+\sin\alpha\cos\beta\,e_2 -\sin\alpha\sin\beta\,e_3\big) ,
    \label{eq:shadow_momentum_decomposition}
\end{align}
where $(\alpha, \beta)$ are the coordinates on the celestial sphere of the observer. Comparing the coordinate components of the two expressions for $p$ gives
\begin{subequations}
\label{eq:shadow_coordinate_angles}
\begin{align}
    \cos\alpha &= \left.\sqrt{\frac{\Sigma}{\Delta_r}}\frac{p^r}{p^0}\right|_{(r_O,y_O)},\\
    \sin\beta &= \left.\frac{1}{\sin\alpha}\sqrt{\frac{\Delta_y}{\Delta_r}}\left(\frac{\sqrt{\Sigma\Delta_r}\,p^\psi}{p^0}-1\right)\right|_{(r_O,y_O)}.
\end{align}
\end{subequations}

\begin{figure*}
    \centering
    \includegraphics[width=0.8\textwidth]{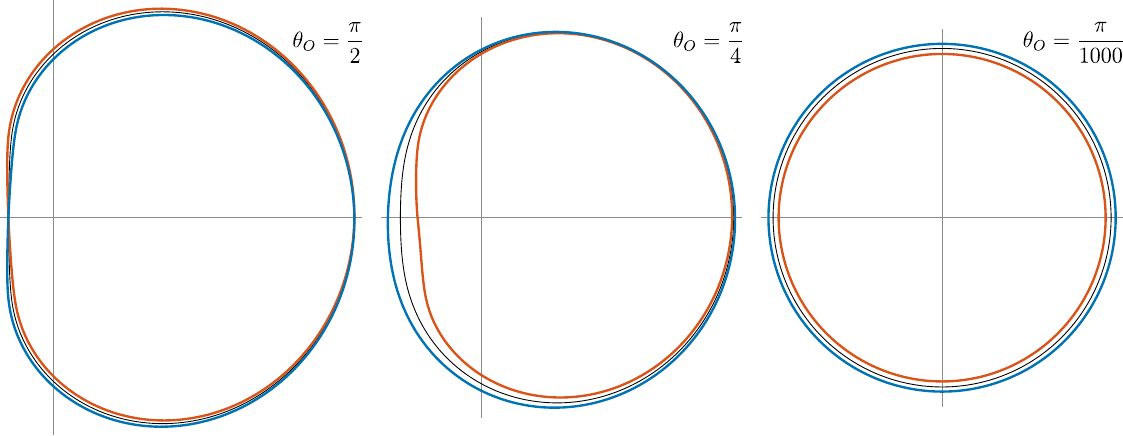}
    \caption{Polarization-dependent Kerr shadows for $a=0.999M$, $r_O=10M$, and $\varepsilon/M=0.5$. From left to right, $\theta_O=\pi/2$, $\pi/4$, and $\pi/1000$. The geodesic shadow is plotted in black, while orange and blue correspond to helicities $s=+1$ and $s=-1$, respectively. The relatively large value of $\varepsilon/M$ is chosen to make the polarization dependence visible.}
    \label{fig:shadow}
\end{figure*}

These relations determine the apparent direction of a spin Hall ray from its momentum at the observer. To obtain the shadow boundary, we must identify the momenta corresponding to rays that asymptote to unstable spherical orbits when traced into the past. We therefore first express the momentum components in terms of the conserved quantities and then determine the critical values of these quantities from the radial trapping conditions.

The conservation laws determine the momentum components algebraically in terms of $E,L,\mathcal D$, as shown in \cref{app:shadow_momentum}. We use the notation $P=r^2E-L$ and $Q=y^2E+L$, so $\mathcal R=P^2+\Delta_r\mathcal D$, with $E>0$, $\mathcal D<0$, and $P>0$ at the observer. We choose the outgoing radial branch appropriate to an observer outside the contributing spherical radii, with $\Delta_r,\Delta_y>0$ and no intervening radial turning point. Substitution into \cref{eq:shadow_coordinate_angles} gives
\begin{subequations}
\label{eq:shadow_momentum_curve}
\begin{align}
    &\sin\left[\beta-\frac{\varepsilon s r}{\Sigma}\left(\frac{Q}{\mathcal D}-\frac{\Delta_r}{P}\right)\right]\bigg|_{(r_O,y_O)} \nonumber\\
    &\quad = \frac{Q+\dfrac{\varepsilon s\sqrt{\mathcal R}}{P}\left(\dfrac{\Delta_y'}{2}-\dfrac{y\Delta_y}{\Sigma}\right)}{\sqrt{\Delta_y\left[-\mathcal D+\dfrac{2\varepsilon s y\sqrt{\mathcal R}}{P\Sigma}(2P+Q)\right]}}\Biggl|_{(r_O,y_O)}, \label{eq:shadow_azimuth}
\end{align}
\begin{align}
    &\cos\alpha = \Bigg\{\frac{\sqrt{\mathcal R}}{P}-\frac{\varepsilon s\Delta_r}{P^2\Sigma}\Bigg[y\left(P+Q+\frac{\mathcal R}{P}\right) \nonumber\\
    &\qquad\qquad\qquad +\frac{r\sqrt{-\Delta_y\mathcal D\mathcal R}}{P}\cos\beta\Bigg]\Bigg\}\Bigg|_{(r_O,y_O)}. \label{eq:shadow_polar_angle}
\end{align}
\end{subequations} 
To select the shadow boundary, we trace the received rays into the past: captured rays reach the horizon, while escaping rays reach an exterior illuminating region. For the nonprincipal branch with $\Delta_r'(r_{\rm p})\ne0$, the first two conditions in \cref{eq:shadow_spherical_conditions} give
\begin{subequations}
\label{eq:shadow_critical_constants}
\begin{align}
    \frac{L}{E} &= r_{\rm p}^2-\frac{4r_{\rm p}\Delta_r(r_{\rm p})}{\Delta_r'(r_{\rm p})},\\
    \frac{\mathcal D}{E^2} &= -\frac{16r_{\rm p}^2\Delta_r(r_{\rm p})}{\Delta_r'(r_{\rm p})^2}.
\end{align}
\end{subequations}
The physical and auxiliary conserved quantities agree through first order, so these are the geodesic expressions. Substituting them into \cref{eq:shadow_momentum_curve} gives the parametrized shadow boundary $(\alpha_\pm(r_{\rm p}),\beta_\pm(r_{\rm p}))$ in terms of the metric functions, the observer's position, the helicity, and the momentum scale.

The parameter interval $r_{\rm p}^{\min} \leq r_{\rm p} \leq r_{\rm p}^{\max}$ is determined by requiring the right-hand side of \cref{eq:shadow_azimuth} to lie in $[-1,1]$. For the numerical prescription used in the figures, its numerator and denominator are evaluated without further expansion in $\varepsilon$. At the endpoints, the right-hand side equals $\pm1$, giving
\begin{align}
    0 &= \Bigg\{\left[Q + \frac{\varepsilon s\sqrt{\mathcal R}}{P}\left(\frac{\Delta_y'}{2} - \frac{y\Delta_y}{\Sigma}\right)\right]^2 \nonumber\\
    &\qquad - \Delta_y\left[-\mathcal D + \frac{2\varepsilon s y\sqrt{\mathcal R}}{P\Sigma}(2 P + Q)\right]\Bigg\}\Bigg|_{(r_O,y_O)}.
    \label{eq:shadow_endpoint}
\end{align}
We retain this squared equation when determining the numerical endpoints; the corresponding Kerr construction is given in \cref{app:shadow_kerr}. Here, the conserved ratios are the functions of $r_{\rm p}$ in \cref{eq:shadow_critical_constants}. Although the critical ratios in \cref{eq:shadow_critical_constants} are unchanged, the endpoints generally shift from the geodesic ones: the spin correction changes which spherical rays can reach the angular position of the observer. We assume a regular black hole exterior with a single smooth shadow boundary determined by radially accessible unstable spherical rays, as is generally the case for nonaccelerating Pleba\'nski--Demia\'nski black holes \cite{Grenzebach2014}.

Unlike the geodesic shadow, the polarization-corrected one depends on the frequency. We fix the momentum normalization at the observer by
\begin{equation}
    \left.p^0\right|_{(r_O,y_O)} = \varepsilon \omega_O = 1.
    \label{eq:shadow_observer_frequency}
\end{equation}
Thus, all rays on a given boundary have the same observed frequency, which is varied by changing the expansion parameter $\varepsilon$. At leading order, this normalization relates the conserved energy to the angular-momentum ratio. Substituting its critical value from \cref{eq:shadow_critical_constants} gives
\begin{subequations} \label{eq:shadow_leading_energy}
\begin{align}
    1 &= \left.\frac{E^{(0)}}{\sqrt{\Sigma \Delta_r}}\left(r^2 - \frac{L^{(0)}}{E^{(0)}}\right)\right|_{(r_O,y_O)}, \\
    E^{(0)} &= \left.\frac{\sqrt{\Sigma \Delta_r}}{r^2 - r_{\rm p}^2 + \dfrac{4 r_{\rm p} \Delta_r(r_{\rm p})}{\Delta_r'(r_{\rm p})}}\right|_{(r_O,y_O)}.
\end{align}
\end{subequations}
The conserved energy therefore varies along the boundary even though the observed frequency is fixed. After substituting the critical ratios, the overall energy scale cancels from the leading angular terms in \cref{eq:shadow_momentum_curve}. Its remaining dependence occurs only in the first-order spin corrections, where the leading energy determined above is sufficient.

This construction assigns celestial directions using momentum. Replacing $p^a$ by the centroid velocity components $v^a=(e^a)_\mu\dot x^\mu$ gives the angle equations in \cref{eq:shadow_velocity_curve}, derived in \cref{app:shadow_velocity}. Their endpoint conditions agree through first order, but give different directions at finite distance. In an asymptotically flat exterior, the two directions agree in the distant-observer limit, with a difference of order $\omega_O^{-1}r_O^{-2}$ at fixed local frequency.

\sect{Kerr shadows}To showcase the polarization-corrected shadows, we consider a subextremal Kerr black hole with $0<a<M$. The metric functions and Boyer--Lindquist polar coordinate are $\Delta_r = r^2 - 2Mr + a^2$, $\Delta_y = a^2 - y^2$, $\Sigma = r^2 + a^2 \cos^2\theta$, and $y = a \cos\theta$. The remaining canonical coordinates are related to Boyer--Lindquist coordinates by $\tau = t_{\rm BL} - a\phi_{\rm BL}$ and $\psi = \frac{\phi_{\rm BL}}{a}$. Defining the spin-dependent Boyer--Lindquist energy and axial angular momentum by $E_{\rm BL} = -\mathcal C_{\partial_{t_{\rm BL}}}$ and $L_z = \mathcal C_{\partial_{\phi_{\rm BL}}}$, the linearity of the Killing charges gives $E = E_{\rm BL}$ and $L = aL_z - a^2E_{\rm BL}$.

In the Schwarzschild limit $a=0$, the spherical-orbit family collapses to the photon sphere at $r_{\rm p}=3M$, and the first-order correction to the celestial angle $\alpha$ vanishes. The shadow therefore remains the same circle as in the geodesic case. Polarization-dependent corrections to $\beta$ can shift individual directions along this circle without changing its boundary.

To display the Kerr shadows, we orthogonally project the celestial sphere onto the plane through its center, which is perpendicular to $e_1$. Taking the horizontal direction along $-e_3$ and the vertical direction along $e_2$ gives
\begin{equation}
    X = -\sin\alpha \sin\beta, \qquad Y = -\sin\alpha \cos\beta.
    \label{eq:shadow_orthogonal_projection}
\end{equation}
These equations are given in \cref{app:shadow_kerr}. 

\Cref{fig:shadow} compares the black hole shadows obtained for $s = \pm 1$, as well as the geodesic shadow, for different $\theta_O$. With the momentum normalization \eqref{eq:shadow_observer_frequency}, $\varepsilon = \omega_O^{-1}$ has dimensions of length. The dimensionless parameter used in the figures is therefore $\varepsilon/M = (M\omega_O)^{-1}$.

\sect{Discussion}We have derived analytic solutions for the gravitational spin Hall equations in Carter spacetimes. The reduction to auxiliary null geodesics allows spin Hall trajectories to be reconstructed algebraically whenever the corresponding null geodesic solutions are known analytically. Independently of these solutions, applying the conservation laws to trapping also yields an analytic description of polarization-dependent corrections to black hole shadows.

We emphasize that the same equations \eqref{eq:spin_hall}, but with $s = \pm 2$, describe the gravitational spin Hall effect of circularly polarized gravitational waves \cite{GSHE_GW}. These analytic spin Hall solutions are therefore particularly relevant to gravitational-wave lensing, where the much longer observed wavelengths can produce substantially larger spin Hall effects than those for electromagnetic radiation in comparable lensing geometries. Avoiding additional integrations of the spin Hall equations can facilitate a systematic exploration of configurations such as hierarchical triple black hole systems, where potentially observable propagation effects have been identified \cite{GSHE_lensing,GSHE_lensing2}. The solutions also provide a basis for comparisons with complementary spin-optical calculations of wave phases, magnifications, and polarized scattering \cite{PhysRevD.109.044027,Frolov2025}. Recent wave-optical calculations provide independent evidence for the gravitational spin Hall effect, revealing helicity-dependent image displacements consistent with the frequency scaling predicted by the spin Hall equations \cite{Nazare2026}. Our analytic solutions offer a basis for quantitative comparisons with such calculations and for assessing the accuracy of the spin Hall approximation.

For the supermassive black holes observed by the Event Horizon Telescope \cite{EHT2019,EHT2022}, the polarization-dependent shadow corrections are far below current observational sensitivity. Nevertheless, the shadow areas provide natural quantities for comparison with black hole absorption cross sections. In geometric optics, the shadow area provides the leading high-frequency absorption cross section. Comparing the polarization-dependent corrections at matched incident frequency and helicity conventions with electromagnetic and gravitational-wave absorption calculations \cite{Dolan2008,leite2017absorption,leite2018absorption} could clarify which finite-frequency effects are described by the spin Hall equations.

\sect{Acknowledgments}This research was funded in whole or in part by the Austrian Science Fund (FWF) \href{https://doi.org/10.55776/PIN9589124}{10.55776/PIN9589124}.

% \clearpage

\appendix
\crefalias{section}{appendix}

\section{Change of observer and reduction to null geodesics}

In this appendix, we establish the behavior of the conserved quantities under the observer and momentum transformations, derive the reduced geodesic equations, and provide the reconstruction formulas for both nonprincipal and principal initial data. We also extend the construction to conformally Carter metrics. Throughout this appendix, equalities are understood through first order in $\varepsilon$, with $\mathcal O(\varepsilon^2)$ remainders omitted unless displayed. First-order corrections may be evaluated on the leading trajectory. The nonprincipal construction assumes nonzero leading $\mathcal D$ on a regular branch, and principal initial data are treated separately in \cref{app:principal_observer}.

\subsection{Invariance of the conserved quantities}
\label{app:observer_invariance}

We show that a change of observer preserves the physical conserved quantities, and that the nonprincipal momentum transformation identifies them with the auxiliary geodesic constants. For the observer change in \cref{eq:observer_centroid_map}, parallel transport all tilded tensors back to $x$ before expanding in $\delta x$. Metric compatibility and the spin constraint give
\begin{subequations}

\begin{align}
    \tilde H &= H,\\
    p \cdot \delta x &= \frac{p_\mu S^{\mu\nu} \tilde t_\nu}{p \cdot \tilde t} = 0.
\end{align}
\end{subequations}
For a conformal Killing vector, the position and spin variations combine as
\begin{equation}
    \tilde{\mathcal C}_\kappa - \mathcal C_\kappa = 2 \delta x^\mu p^\nu \nabla_{(\mu}\kappa_{\nu)} = \frac12 (p \cdot \delta x) \nabla_\mu \kappa^\mu = 0.
\end{equation}
This proves the invariance of $E$ and $L$.

For the hidden conservation law, the conformal Killing--Yano equation reads
\begin{align}
    \nabla_\rho h_{\mu\nu} &= \frac13 \left(g_{\rho\mu} \nabla^\alpha h_{\alpha\nu} - g_{\rho\nu} \nabla^\alpha h_{\alpha\mu}\right) + \nabla_{[\rho}h_{\mu\nu]}.
\end{align}
Substitution into the definitions of $\mathcal K$ and $\mathcal L$ in \cref{eq:D_CKY} cancels all totally antisymmetric derivatives and gives
\begin{equation}
    \nabla_\rho \mathcal K_{\mu\nu} - 2 \mathcal L_{\rho(\mu\nu)} = \frac23 \big(\nabla^\sigma h_\sigma{}^\alpha\big) h_{\alpha(\mu} g_{\nu)\rho}.
\end{equation}
This is Eq.~(4.14b) of Ref.~\cite{GSHE_int} adapted to our definition of $\mathcal L$: the metric term $-\frac23 (\nabla^\sigma h_\sigma{}^\rho) h_{\rho[\alpha} g_{\beta]\mu}$ retained in that reference is omitted here because its contraction with $S^{\alpha\beta} p^\mu$ vanishes. The displacement of $\mathcal L$ contributes only at second order, so
\begin{align}
    \tilde{\mathcal D} - \mathcal D &= \delta x^\rho p^\mu p^\nu \left(\nabla_\rho \mathcal K_{\mu\nu} - 2 \mathcal L_{\rho(\mu\nu)}\right) \nonumber\\
    &= \frac23 (p \cdot \delta x) \big(\nabla^\sigma h_\sigma{}^\alpha\big) h_{\alpha\mu} p^\mu = 0.
\end{align}
Thus all four quantities are preserved by any regular observer change, including at principal null directions. Closure of $h$ is not required.

For the nonprincipal momentum transformation in \cref{eq:carter_auxiliary_momentum}, work at $\tilde x = \hat x$. The Carter tensors obey
\begin{subequations}

\begin{align}
    h^\mu{}_\rho f^\rho{}_\nu &= ry\,\delta^\mu_\nu,\\
    k_\mu f^\mu{}_\nu \tilde p^\nu &= \tilde p^\alpha \nabla_\alpha(ry).
\end{align}
\end{subequations}
The second identity follows by differentiating the first along a leading null geodesic and using the conformal Killing--Yano and Killing--Yano equations. For either $\kappa = k,m$, contracting the vanishing Lie derivative $[\kappa,h] = 0$ with $\tilde p^\mu f^\nu{}_\rho \tilde p^\rho$ gives
\begin{equation}
    h^\mu{}_\alpha f^\nu{}_\beta \tilde p^\alpha \tilde p^\beta \nabla_\mu \kappa_\nu = \tilde p^\nu \Big[(k \cdot \tilde p) \kappa_\mu f^\mu{}_\nu - (\kappa \cdot \tilde p) \nabla_\nu(ry)\Big].
\end{equation}
Inserting the adapted spin tensor from \cref{eq:adapted_spin} and using $k \cdot \tilde p = -E$ at leading order yields
\begin{align}
    \frac12 \tilde S^{\mu\nu} \nabla_\mu \kappa_\nu &= \frac{\varepsilon s}{\mathcal D} \tilde p^\nu \Big[E \kappa_\mu f^\mu{}_\nu + (\kappa \cdot \tilde p) \nabla_\nu(ry)\Big] \nonumber\\
    &= \kappa_\mu (\hat p^\mu - \tilde p^\mu).
\end{align}
The momentum correction therefore reproduces the spin contribution to each Killing charge. For $\kappa = k$, this contribution vanishes, so $E = -k \cdot \tilde p = -\hat p_\tau$. For $\kappa = m$, the contribution is generally nonzero:
\begin{align}
    L = \tilde p_\psi + \frac{\varepsilon s}{\mathcal D} \tilde p^\nu \Big[E m_\mu f^\mu{}_\nu + L \nabla_\nu(ry)\Big] = \hat p_\psi.
\end{align}

For the hidden quantity, the principal-tensor algebra gives
\begin{subequations}

\begin{align}
    \mathcal K_{\mu\nu} \tilde p^\mu f^\nu{}_\rho \tilde p^\rho &= 0,\\
    \tilde S^{\mu\nu} h_{\mu\nu} &= 0,\\
    \tilde S^\mu{}_\nu h^\nu{}_\rho \tilde p^\rho &= -\varepsilon s f^\mu{}_\rho \tilde p^\rho.
\end{align}
\end{subequations}
Since $h$ is closed in the Carter metric, its spin coefficient reduces to Eq.~(4.3) of Ref.~\cite{GSHE_int} with the same metric term omitted. Contracting with the adapted spin tensor then gives
\begin{subequations}
\begin{align}
    \mathcal L_{\alpha\beta\mu} &= -2 h_{\alpha\beta} k_\mu - 2 k_{[\alpha} h_{\beta]\mu},\\
    \mathcal L_{\alpha\beta\mu} \tilde S^{\alpha\beta} \tilde p^\mu &= 2 \varepsilon s\,\tilde p^\alpha \nabla_\alpha(ry).
\end{align}
\end{subequations}
Using $\mathcal K_{\mu\nu} \tilde p^\mu \tilde p^\nu = \mathcal D + \mathcal O(\varepsilon)$ inside the momentum correction now gives
\begin{align}
    \hat{\mathcal D} &= \mathcal K_{\mu\nu} \hat p^\mu \hat p^\nu = \mathcal K_{\mu\nu} \tilde p^\mu \tilde p^\nu + 2 \varepsilon s\,\tilde p^\alpha \nabla_\alpha(ry) = \mathcal D.
\end{align}
Finally, antisymmetry of $f$ and the null constraint imply $\hat p_\mu\hat p^\mu=\tilde p_\mu\tilde p^\mu=0$. Thus
\begin{equation}
    (\hat H, \hat E, \hat L, \hat{\mathcal D}) = (H, E, L, \mathcal D) + \mathcal O(\varepsilon^2),
\end{equation}
where the hatted quantities are the constants of the auxiliary geodesic.

\subsection{Transformed equations and reduction to null geodesics}
\label{app:shift_dynamics}

We derive the shifted spin and equations of motion in \cref{eq:adapted_spin,eq:adapted_gshe}, then verify that the momentum correction and parameter change give the null geodesic equations. We first work at the original centroid, writing $(hp)^\mu = h^\mu{}_\nu p^\nu$ and $(fp)^\mu = f^\mu{}_\nu p^\nu$.

The observer in \cref{eq:adapted_observer} is timelike on the nonprincipal branch because $\tilde t \cdot \tilde t = -[(p^2)^2 + (p^3)^2] < 0$ at leading order. In the Carter frame,
\begin{subequations}
\label{eq:app_principal_vectors}
\begin{align}
    (hp)^a &= (-r p^1, -r p^0, y p^3, -y p^2),\\
    (fp)^a &= (-y p^1, -y p^0, -r p^3, r p^2).
\end{align}
\end{subequations}
Both vectors are orthogonal to $p$ and $\tilde t$, and the null constraint gives
\begin{subequations}
\begin{align}
    (hp) \cdot (hp) &= (fp) \cdot (fp) = -\mathcal D + \mathcal O(\varepsilon),\\
    (hp) \cdot (fp) &= 0.
\end{align}
\end{subequations}
Substituting \cref{eq:adapted_observer,eq:spin_tensor} into the displacement formula \eqref{eq:observer_shift_vector} gives
\begin{equation}
    \delta x^\mu = \frac{\varepsilon s p^1}{\mathcal D p^0} \big[r (fp)^\mu - y (hp)^\mu\big].
\end{equation}
This reproduces \cref{eq:carter_displacement}. The displacement is orthogonal to $t$ and $p$, and its radial component vanishes. Inserting it into the spin transformation in \cref{eq:observer_centroid_map} gives
\begin{equation}
    S^{\mu\nu} + 2 p^{[\mu} \delta x^{\nu]} = -\frac{2 \varepsilon s}{\mathcal D} (hp)^{[\mu} (fp)^{\nu]}.
    \label{eq:app_spin_displacement_identity}
\end{equation}
Parallel transport to the displaced point therefore gives the shifted spin tensor \eqref{eq:adapted_spin}.

The closed conformal Killing--Yano and Killing--Yano equations imply, along the leading null geodesic,
\begin{subequations}
\label{eq:app_principal_transport}
\begin{align}
    \frac{D(hp)^\mu}{\dd\lambda} &= (k \cdot p) p^\mu = -E p^\mu,\\
    \frac{D(fp)^\mu}{\dd\lambda} &= 0.
\end{align}
\end{subequations}
Differentiating the shifted spin and using the Mathisson--Papapetrou--Dixon spin equation therefore gives
\begin{equation}
    \frac{\tilde D \tilde S^{\mu\nu}}{\dd\lambda} = \frac{2 \varepsilon s E}{\mathcal D} \tilde p^{[\mu} (f\tilde p)^{\nu]} = 2 \tilde p^{[\mu} \frac{\dd\tilde x^{\nu]}}{\dd\lambda}.
\end{equation}
This fixes the shifted velocity up to a multiple of $\tilde p$. To determine that multiple, pull the velocity back to the original point. The original spin supplementary condition gives $t \cdot \dot x = t \cdot p = -p^0$. Differentiating $t \cdot \delta x = 0$ and using \cref{eq:observer_shifted_velocity} then gives
\begin{equation}
    t \cdot G^{-1}\dot{\tilde x} = -p^0 - \delta x \cdot \nabla_p t.
\end{equation}
Using $t \cdot (hp) = r p^1$, $t \cdot (fp) = y p^1$, and \cref{eq:app_principal_transport}, the required contraction is
\begin{align}
    \big[r(fp) - y(hp)\big] \cdot \nabla_p t &= r \frac{\dd(y p^1)}{\dd\lambda} - y \frac{\dd(r p^1)}{\dd\lambda} + E y p^0 \nonumber\\
    &= p^1(r \dot y - y \dot r) + E y p^0.
\end{align}
Here $\dot r = p^\alpha \nabla_\alpha r$ and $\dot y = p^\alpha \nabla_\alpha y$ are needed only at leading order. Combining the transverse and longitudinal parts gives
\begin{align}
    G^{-1}\dot{\tilde x} &= p + \frac{\varepsilon s E}{\mathcal D} (fp) \nonumber\\
    &\quad + \frac{\varepsilon s}{\mathcal D} \left[\left(\frac{p^1}{p^0}\right)^2 (r \dot y - y \dot r) + \frac{2 E y p^1}{p^0}\right] p.
    \label{eq:app_expanded_shifted_velocity}
\end{align}
This is the expanded shifted velocity underlying \cref{eq:adapted_velocity}. After deriving the momentum equation, we identify its factorization in terms of $\hat p$ and the parameter factor $N$ below.

For the momentum equation, our coordinate curvature convention is
\begin{equation}
    [\nabla_\alpha, \nabla_\beta] V^\mu = R^\mu{}_{\nu\alpha\beta} V^\nu
\end{equation}
for any vector field $V$. Combining the parallel-propagator derivative in \cref{eq:observer_shifted_force} with \cref{eq:app_spin_displacement_identity} gives
\begin{equation}
    G^\mu{}_{\nu'} \frac{\tilde D \tilde p^{\nu'}}{\dd\lambda} = \frac{\varepsilon s}{\mathcal D} R^\mu{}_{\nu\alpha\beta} p^\nu (hp)^\alpha (fp)^\beta.
    \label{eq:app_transformed_force}
\end{equation}
Here $G$ transports back to the original point. The required Carter curvature identity is
\begin{equation}
    R^\mu{}_{\nu\alpha\beta} p^\nu (hp)^\alpha (fp)^\beta = -\left[p^\alpha p^\beta \nabla_\alpha \nabla_\beta(ry)\right] p^\mu.
    \label{eq:app_principal_curvature_identity}
\end{equation}
To verify it without imposing field equations, the relevant principal-frame components are
\begin{subequations}
\label{eq:app_mixed_curvature}
\begin{align}
    R_{0202} &= R_{0303} = -R_{1212} = -R_{1313},\\
    R_{0123} &= 2 R_{0213} = -2 R_{0312},\\
    R_{0213} &= \frac{y \Delta_r' - r \Delta_y'}{2 \Sigma^2} + \frac{2 ry(\Delta_y - \Delta_r)}{\Sigma^3}.
\end{align}
\end{subequations}
Apart from diagonal bivector components, the curvature couples only $(01,23)$, $(02,13)$, and $(03,12)$. Substituting \cref{eq:app_principal_vectors} into the contraction and differentiating $ry$ directly gives
\begin{subequations}
\begin{align}
    R^a{}_{bcd} p^b (hp)^c (fp)^d &= \Sigma R_{0213} \big[(p^2)^2 + (p^3)^2\big] p^a,\\
    p^\alpha p^\beta \nabla_\alpha \nabla_\beta(ry) &= -\Sigma R_{0213} \big[(p^2)^2 + (p^3)^2\big].
\end{align}
\end{subequations}
The null constraint was used in both equations. Their comparison proves \cref{eq:app_principal_curvature_identity}, and substitution into \cref{eq:app_transformed_force} gives the shifted force \eqref{eq:adapted_force}.

Now define $\hat x = \tilde x$ and $\hat p$ by \cref{eq:carter_auxiliary_momentum}. Since $f\tilde p$ is parallel transported at leading order,
\begin{equation}
    \frac{\tilde D(\hat p^\mu - \tilde p^\mu)}{\dd\lambda} = \frac{\varepsilon s}{\mathcal D} \left[\tilde p^\alpha \tilde p^\beta \nabla_\alpha \nabla_\beta(ry)\right] \tilde p^\mu.
\end{equation}
This cancels the shifted force. Subtracting the same momentum correction from \cref{eq:app_expanded_shifted_velocity} gives
\begin{align}
    \frac{\dd\hat x^\mu}{\dd\lambda} &= \Bigg\{1 + \frac{\varepsilon s}{\mathcal D} \Bigg[\left(\frac{p^1}{p^0}\right)^2 (r \dot y - y \dot r) \nonumber\\
    &\qquad + \frac{2 E y p^1}{p^0} - \frac{\dd(ry)}{\dd\lambda}\Bigg]\Bigg\} \hat p^\mu.
\end{align}
The correction in braces is evaluated at the original point. With $Q = y^2 E + L$, the leading relations are
\begin{subequations}
\begin{align}
    \mathcal D &= -\Sigma \big[(p^0)^2 - (p^1)^2\big],\\
    \delta y &= \frac{\varepsilon s p^1 Q}{\mathcal D p^0},\\
    \frac{Q}{\Sigma} &= E - \sqrt{\frac{\Delta_r}{\Sigma}}\,p^0,\\
    \dot r &= \sqrt{\frac{\Delta_r}{\Sigma}}\,p^1.
\end{align}
\end{subequations}
They rewrite the braces as
\begin{equation}
    \frac{\Sigma(\tilde x)}{\Sigma(x)} \left[1 - \frac{\varepsilon s\,p^\alpha(y \nabla_\alpha r - r \nabla_\alpha y)}{\Sigma(x)(p \cdot t)^2}\right].
\end{equation}
To express the parameter factor entirely at the shifted point, use
\begin{equation}
    \frac{1}{\Sigma(x)} = \frac{1}{\Sigma(\tilde x)} + \frac{\delta x^\mu \nabla_\mu \Sigma}{\Sigma(\tilde x)^2}.
\end{equation}
The displacement in the correction is parallel transported to $\tilde x$. All other first-order factors may likewise be evaluated there, with $p$ replaced by $\tilde p$. This gives $N$ in \cref{eq:carter_parameter_factor} and proves
\begin{subequations}

\begin{align}
    \frac{\dd\hat x^\mu}{\dd\lambda} &= \Sigma(\hat x) N \hat p^\mu,\\
    \frac{\hat D \hat p^\mu}{\dd\lambda} &= 0.
\end{align}
\end{subequations}
Consequently, $\dd\gamma = N \dd\lambda$ gives \cref{eq:carter_auxiliary_geodesic}. An affine geodesic parameter satisfies $\dd\hat\lambda = \Sigma(\hat x) \dd\gamma$.

\subsection{Reconstruction for nonprincipal null directions}

We now recover the Carter-observer spin Hall position, physical momentum, and spin from an auxiliary geodesic, and give the optional conversion back to the original parameter. Every first-order correction in the inverse map can be evaluated on the auxiliary geodesic at the same value of $\gamma$.

In the original variables, $p^1/p^0 = \Delta_r p_r/P + \mathcal O(\varepsilon)$, where $P = r^2 E - L$ and $Q = y^2 E + L$. Thus the displacement \cref{eq:carter_displacement} has coordinate components
\begin{equation}
    (\delta r, \delta y, \delta\psi, \delta\tau) = \frac{\varepsilon s \Delta_r p_r}{\mathcal D P} (0, Q, -p_y, y^2 p_y).
\end{equation}
Reversing this displacement and evaluating its coefficients on the auxiliary geodesic gives the positions in \cref{eq:carter_inverse_coordinates}. The full covariant inverse map is
\begin{subequations}

\begin{align}
    x &= \exp_{\hat x}(-\delta x),\\
    p^\mu &= G^\mu{}_{\nu'} \Bigg[\hat p^{\nu'} - \frac{\varepsilon s}{\hat{\mathcal D}} \hat p^{\rho'} \Big(\hat E f^{\nu'}{}_{\rho'}  + \nabla_{\rho'}(ry)\,\hat p^{\nu'}\Big)\Bigg].
\end{align}
\end{subequations}
Here $G$ transports from $\hat x$ to the reconstructed point, and all quantities inside the brackets belong to $\hat x$. Expanding this parallel transport in a common coordinate chart gives the physical covector as
\begin{align}
    p_\mu &= \hat p_\mu - \frac{\varepsilon s}{\hat{\mathcal D}} \hat p^\nu \left[\hat E f_{\mu\nu} + \nabla_\nu(ry)\,\hat p_\mu\right] \nonumber\\
    &\quad - \Gamma^\rho_{\mu\nu}(\hat x) \hat p_\rho \delta x^\nu.
    \label{eq:app_coordinate_momentum_shift}
\end{align}
The last term accounts for the displacement of the point at which the momentum is represented.

The auxiliary radial and polar covector components are given by \cref{eq:carter_auxiliary_covector_gamma}, and $\hat p_\tau = -\hat E$, $\hat p_\psi = \hat L$. Substituting the Carter metric and displacement into \cref{eq:app_coordinate_momentum_shift} yields
\begin{subequations}
\label{eq:app_inverse_covector_components}
\begin{align}
    p_\tau &= -\hat E + \frac{\varepsilon s \Delta_r \hat p_r}{P \Sigma^2} \left[\hat y(\Delta_y - \Delta_r) - \frac{\Sigma \Delta_y'}{2}\right],\\
    p_\psi &= \hat L - \frac{\varepsilon s \hat r \Delta_r \Delta_y}{P \Sigma} \hat p_y \nonumber\\
    &\quad - \frac{\varepsilon s \Delta_r \hat p_r}{P \Sigma^2} \left[\hat y \hat r^2 \Delta_y + \hat y^3 \Delta_r - \frac{\hat r^2 \Sigma \Delta_y'}{2}\right],\\
    p_r &= \hat p_r - \frac{\varepsilon s \hat E \hat y}{P},\\
    p_y &= \hat p_y + \frac{\varepsilon s \hat r(\Delta_r Q + P \Delta_y)}{P \Sigma \Delta_y} \nonumber\\
    &\quad - \frac{\varepsilon s \Delta_r \hat p_r}{\hat{\mathcal D} P} \left(2 \hat E \hat y - \frac{Q \Delta_y'}{\Delta_y}\right) \hat p_y.
\end{align}
\end{subequations}
All functions on the right are evaluated at the auxiliary coordinates, with $P = \hat r^2 \hat E - \hat L$, $Q = \hat y^2 \hat E + \hat L$, and $\Sigma = \hat r^2 + \hat y^2$. The left-hand sides belong to the reconstructed physical point. Primes denote derivatives of the corresponding metric function. The time and azimuthal components also follow from the spin-corrected Killing charges. The spin tensor is then given by \cref{eq:spin_tensor} at the reconstructed point. No inverse radial or polar momentum occurs, so the reconstruction remains regular at ordinary turning points on the nonprincipal branch.

Conversely, physical initial data for the spin Hall equations determine the auxiliary null geodesic initial point by the forward displacement and its momentum by \cref{eq:carter_auxiliary_momentum} after parallel transport. The complete physical phase-space trajectory then follows from the auxiliary geodesic by the algebraic operations above. Only recovery of the original parametrization requires an additional quadrature: inversion of \cref{eq:carter_parameter_factor} gives
\begin{align}
    \frac{\dd\lambda}{\dd\gamma} = \frac{1}{N} &= \Sigma - \frac{2 \varepsilon s \hat y Q \Delta_r \hat p_r}{\hat{\mathcal D} P} \nonumber\\
    &\quad + \frac{\varepsilon s \Delta_r}{P^2} \left(\hat y \Delta_r \hat p_r - \hat r \Delta_y \hat p_y\right).
    \label{eq:app_reconstruct_parameter}
\end{align}
Every quantity on the right is evaluated on the auxiliary geodesic. The negative displacement term comes from evaluating $\Sigma(x)$ at the reconstructed point. Integrating this equation and inverting $\lambda(\gamma)$ gives the solution in the original spin Hall parameter $\lambda$ when needed.

\subsection{Principal null initial data}
\label{app:principal_observer}

We first construct a regular auxiliary geodesic for exactly principal initial momentum and derive the corresponding reconstruction formulas. A subscript zero denotes evaluation at the initial point.

The observer in \cref{eq:adapted_observer} is null in this case, so we instead parallel transport a timelike observer along the leading principal geodesic. We normalize the auxiliary affine momentum to equal the physical momentum initially, with $\hat E = -p_\tau|_0$. The leading ray has constant $y = y_0$, radial orientation $\chi = \pm 1$, and we write $\Sigma_0=r_0^2+y_0^2>0$. Its momentum and radial evolution satisfy
\begin{subequations}

\begin{align}
    p &= p^0(e_0 + \chi e_1),\\
    \frac{\dd r}{\dd\lambda} &= \chi \hat E,
\end{align}
\end{subequations}
where $p^0 = \hat E \sqrt{\Sigma/\Delta_r}$. All first-order terms below are evaluated on this ray, with $\Delta_y$ and $\Delta_y'$ evaluated at $y_0$. The spin tensor has only the independent component $S^{23} = \varepsilon s \chi$, and the Carter-observer equations become
\begin{subequations}

\begin{align}
    \dot x &= p + \frac{\varepsilon s \sqrt{\Delta_y}}{\Sigma^{3/2}}(-r e_2 + \chi y_0 e_3),\\
    \frac{Dp^\mu}{\dd\lambda} &= 2 \varepsilon s R_{0213} p^\mu.
\end{align}
\end{subequations}
The curvature component is given in \cref{eq:app_mixed_curvature}. Although the force is longitudinal, the anomalous velocity makes the physical momentum depart from the principal direction relative to the Carter frame.

Choose the auxiliary observer to satisfy $\tilde t|_0 = e_0|_0$ and define the displacement by
\begin{subequations}

\begin{align}
    p^\alpha \nabla_\alpha \tilde t^\mu &= 0,\\
    \delta x^\mu &= \frac{S^{\mu\nu} \tilde t_\nu}{p \cdot \tilde t}.
\end{align}
\end{subequations}
The position, momentum, and spin transform by \cref{eq:observer_centroid_map}. In particular, the shifted spin is the parallel transport of $S^{\mu\nu} + 2 p^{[\mu} \delta x^{\nu]}$. To determine $\delta x$ explicitly, use the leading screen transport equations
\begin{subequations}
\label{eq:app_principal_screen_transport}
\begin{align}
    \nabla_p e_2 &= \frac{y_0 \sqrt{\Delta_y}}{\Sigma^{3/2}} p + \frac{\hat E y_0}{\Sigma} e_3,\\
    \nabla_p e_3 &= \frac{\chi r \sqrt{\Delta_y}}{\Sigma^{3/2}} p - \frac{\hat E y_0}{\Sigma} e_2.
\end{align}
\end{subequations}
Writing $\delta x = \delta x^2 e_2 + \delta x^3 e_3$, cancellation of the transverse anomalous velocity requires
\begin{subequations}

\begin{align}
    \chi \hat E \partial_r \delta x^2 - \frac{\hat E y_0}{\Sigma} \delta x^3 &= \frac{\varepsilon s r \sqrt{\Delta_y}}{\Sigma^{3/2}},\\
    \chi \hat E \partial_r \delta x^3 + \frac{\hat E y_0}{\Sigma} \delta x^2 &= -\frac{\varepsilon s \chi y_0 \sqrt{\Delta_y}}{\Sigma^{3/2}}.
\end{align}
\end{subequations}
The solution with $\delta x|_0 = 0$ is
\begin{equation}
    \delta x = \frac{\varepsilon s \sqrt{\Delta_y}(r - r_0)}{\hat E \Sigma_0 \sqrt{\Sigma}}(\chi r_0 e_2 - y_0 e_3).
\end{equation}
Consequently, $\delta r = 0$ and the remaining coordinate components are
\begin{subequations}
\label{eq:app_principal_coordinate_shift}
\begin{align}
    \delta y &= \frac{\varepsilon s \chi \Delta_y r_0(r - r_0)}{\hat E \Sigma_0 \Sigma},\\
    \delta\psi &= \frac{\varepsilon s y_0(r - r_0)}{\hat E \Sigma_0 \Sigma},\\
    \delta\tau &= -y_0^2 \delta\psi.
\end{align}
\end{subequations}
This displacement is finite on a regular principal ray, and reversing its coordinate components gives \cref{eq:principal_inverse_coordinates}.

The longitudinal part of \cref{eq:app_principal_screen_transport} gives the shifted velocity,
\begin{equation}
    \frac{\dd\tilde x^\mu}{\dd\lambda} = \left[1 - \frac{\varepsilon s \chi y_0 \Delta_y(r - r_0)^2}{\hat E \Sigma_0 \Sigma^2}\right] \tilde p^\mu.
\end{equation}
The principal curvature identities, in coordinate components, are
\begin{subequations}

\begin{align}
    R^\mu{}_{\nu\alpha\beta} p^\nu (e_2)^\alpha p^\beta &= 0,\\
    R^\mu{}_{\nu\alpha\beta} p^\nu (e_3)^\alpha p^\beta &= 0.
\end{align}
\end{subequations}
Since $\delta x$ lies in the span of $e_2,e_3$, these identities remove the displacement contribution in \cref{eq:observer_shifted_force}, leaving
\begin{equation}
    \frac{\tilde D \tilde p^\mu}{\dd\lambda} = 2 \varepsilon s R_{0213} \tilde p^\mu.
\end{equation}
Since $k = \partial_\tau$ is Killing and the shifted velocity is parallel to $\tilde p$, its contraction obeys $\dd\tilde p_\tau/\dd\lambda = 2 \varepsilon s R_{0213} \tilde p_\tau$. Keeping $\hat x = \tilde x$, the force is therefore eliminated by
\begin{subequations}

\begin{align}
    \hat p^\mu &= -\frac{\hat E}{\tilde p_\tau} \tilde p^\mu,\\
    \frac{\hat D \hat p^\mu}{\dd\lambda} &= 0.
\end{align}
\end{subequations}
An affine parameter and the separation parameter are then determined by
\begin{subequations}
\begin{align}
    \frac{\dd\hat\lambda}{\dd\lambda} &= -\frac{\tilde p_\tau}{\hat E} \left[1 - \frac{\varepsilon s \chi y_0 \Delta_y(r - r_0)^2}{\hat E \Sigma_0 \Sigma^2}\right], \label{eq:app_principal_affine_clock}\\
    \dd\gamma &= \frac{\dd\hat\lambda}{\Sigma(\hat x)}. \label{eq:app_principal_gamma}
\end{align}
\end{subequations}
Initially, $\delta x = 0$ and $\tilde p_\tau = -\hat E$, so $\hat p = p$ and $\dd\hat\lambda/\dd\lambda = 1$.

Evaluating the physical conserved quantities on the principal initial data fixes their relation to the auxiliary geodesic constants:
\begin{subequations}
\label{eq:app_principal_cons}
\begin{align}
    \mathcal D &= 4 \varepsilon s \chi y_0 E, \\
    \hat{\mathcal D} &= 0,\\
    \hat E &= E + \varepsilon s \chi \left[\frac{y_0(\Delta_r - \Delta_y)}{\Sigma^2} + \frac{\Delta_y'}{2 \Sigma}\right]_{r = r_0}, \label{eq:app_principal_energy_normalization}\\
    L &= -E y_0^2 + \varepsilon s \chi \left(\frac{y_0 \Delta_y}{\Sigma_0} - \frac{\Delta_y'}2\right), \\
    \hat L &= -\hat E y_0^2.
\end{align}
\end{subequations}
Thus the full physical constants need not equal the auxiliary geodesic constants on this branch. In particular, principal leading momentum does not imply $\mathcal D = 0$ through first order. The auxiliary geodesic has $\hat y = y_0$ and $\hat p_y = 0$, while its remaining coordinates satisfy
\begin{subequations}

\begin{align}
    \frac{\dd\hat r}{\dd\gamma} &= \chi \hat E \Sigma,\\
    \frac{\dd\hat\psi}{\dd\gamma} &= \frac{\hat E \Sigma}{\Delta_r},\\
    \frac{\dd\hat\tau}{\dd\gamma} &= \frac{\hat E \hat r^2 \Sigma}{\Delta_r}.
\end{align}
\end{subequations}
The nonzero covector components are
\begin{subequations}
\begin{align}
    \hat p_\tau &= -\hat E,\\
    \hat p_\psi &= -\hat E y_0^2,\\
    \hat p_r &= \frac{\chi \hat E \Sigma}{\Delta_r}.
\end{align}
\end{subequations}
Here and below, background quantities on the right are evaluated at $(\hat r,y_0)$.

To recover the physical momentum, reverse the rescaling and parallel transport. In components,
\begin{equation}
    p_\mu = -\frac{\tilde p_\tau}{\hat E} \hat p_\mu - \Gamma^\nu_{\mu\alpha} \hat p_\nu \delta x^\alpha.
    \label{eq:app_principal_inverse_covariant_components}
\end{equation}
The connection term vanishes for $\mu = \tau$, so $\tilde p_\tau = p_\tau$. The energy charge determines this component directly, and the remaining connection terms give
\begin{subequations}
\label{eq:app_principal_inverse_momentum}
\begin{align}
    p_\tau &= -E - \varepsilon s \chi \left[\frac{y_0(\Delta_r - \Delta_y)}{\Sigma^2} + \frac{\Delta_y'}{2 \Sigma}\right],\\
    p_\psi &= y_0^2 p_\tau + \frac{\varepsilon s \chi \Delta_y y_0(\hat r^2 - r_0^2)}{\Sigma_0 \Sigma},\\
    p_r &= -\frac{\chi}{\Delta_r}(\hat r^2 p_\tau + p_\psi),\\
    p_y &= \frac{\varepsilon s(\hat r - r_0)(\hat r r_0 + y_0^2)}{\Sigma_0 \Sigma}.
\end{align}
\end{subequations}
The physical components belong to the reconstructed centroid and are evaluated successively in the displayed order. Their differences from the auxiliary components vanish initially. Together with \cref{eq:principal_inverse_coordinates}, these formulas reconstruct the complete physical phase-space trajectory algebraically; the spin follows from \cref{eq:spin_tensor}. Substituting $\tilde p_\tau = p_\tau$ into \cref{eq:app_principal_affine_clock} also makes the parameter conversion explicit, if the original parameter is required.

The same construction also applies when the initial momentum is principal only at leading order, with transverse frame components $p^2|_0,p^3|_0=\mathcal O(\varepsilon)$ and $p^1|_0=\chi p^0|_0+\mathcal O(\varepsilon^2)$. The observer transport and displacement are still evaluated on the leading principal ray, but the auxiliary null geodesic is chosen to have the full physical initial data:
\begin{align}
    \hat x|_0 &= x|_0, \qquad \hat p_\mu|_0 = p_\mu|_0,\\
    \hat E &= -p_\tau|_0, \qquad \hat L = p_\psi|_0, \qquad \hat{\mathcal D} = \left.\mathcal K^{\mu\nu}p_\mu p_\nu\right|_0.
\end{align}
This geodesic is determined by \cref{eq:carter_separated}, without imposing the exactly principal restrictions $\hat{\mathcal D}=0$ and $\hat L=-\hat E y_0^2$. On a fixed regular interval, it satisfies $\hat y-y_0=\mathcal O(\varepsilon)$, $\hat p_y=\mathcal O(\varepsilon)$, $\hat L+\hat E y_0^2=\mathcal O(\varepsilon)$, and $\hat{\mathcal D}=\mathcal O(\varepsilon^2)$. The latter must nevertheless be retained when solving the auxiliary geodesic equations: both terms in the polar potential are of order $\varepsilon^2$, and their square root determines the transverse motion at order $\varepsilon$.

The displacement in \cref{eq:app_principal_coordinate_shift}, evaluated at $r=\hat r$ with $y_0$ fixed, reconstructs the position as
\begin{equation}
    x^\mu(\gamma) = \hat x^\mu(\gamma) - \delta x^\mu(\gamma).
\end{equation}
Thus the leading $y_0$ in the second equation of \cref{eq:principal_inverse_coordinates} is replaced by $\hat y(\gamma)$, while every explicit first-order correction is unchanged. The connection term in \cref{eq:app_principal_inverse_covariant_components} still vanishes for $\mu=\tau$ through first order, so $\tilde p_\tau=p_\tau$, with
\begin{equation}
    p_\tau = -E - \varepsilon s\chi\left[\frac{y_0[\Delta_r(\hat r)-\Delta_y(y_0)]}{(\hat r^2+y_0^2)^2} + \frac{\Delta_y'(y_0)}{2(\hat r^2+y_0^2)}\right].
\end{equation}
All momentum components then follow from \cref{eq:app_principal_inverse_covariant_components}, using the full auxiliary momentum and $\tilde p_\tau=p_\tau$. The energy normalization \eqref{eq:app_principal_energy_normalization} and parameter conversion \eqref{eq:app_principal_affine_clock} remain valid. In \cref{eq:app_principal_gamma}, $\Sigma(\hat x)$ is evaluated on the full auxiliary geodesic. The explicit principal geodesic components and the remaining expanded momentum formulas above apply when the transverse initial momentum vanishes exactly.

\subsection{Extension to conformally Carter metrics}
\label{app:conformal_carter}

We extend the Carter solutions to a conformally related metric by deriving the transformations of the momentum, spin, and parameter, and verifying the conserved quantities. Let
\begin{equation}
    \bar g_{\mu\nu} = \Omega^2 g_{\mu\nu},
\end{equation}
where $g$ is the Carter metric and $\Omega = \Omega(r,y)$ is smooth and positive. The required map, evaluated at the same spacetime point, is
\begin{subequations}
\label{eq:conformal_ray_map}
\begin{align}
    \bar x^\mu &= x^\mu,\\
    \bar t^\mu &= \Omega^{-1} t^\mu,\\
    \bar p_\mu &= p_\mu + S_\mu{}^\nu \nabla_\nu \log\Omega,\\
    \bar S^{\mu\nu} &= \Omega^{-2} S^{\mu\nu},\\
    \dd\bar\lambda &= \Omega^2 \dd\lambda.
\end{align}
\end{subequations}
The constraints $S^{\mu\nu}t_\nu=S^{\mu\nu}p_\nu=p_\mu p^\mu=0$ imply the corresponding barred constraints, and the helicity relation \eqref{eq:spin_tensor} gives the displayed spin transformation. The parameter normalization is also preserved:
\begin{equation}
    \frac{\dd x^\mu}{\dd\bar\lambda} \bar t_\mu = \Omega^{-1} p \cdot t = \bar p \cdot \bar t.
\end{equation}
For the position equation, the conformal connection gives
\begin{equation}
    \bar\nabla_\alpha \bar t_\beta = \Omega \Big(\nabla_\alpha t_\beta - t_\alpha \nabla_\beta \log\Omega + g_{\alpha\beta} t^\rho \nabla_\rho \log\Omega\Big),
\end{equation}
and hence
\begin{equation}
    \frac{\bar S^{\mu\nu} \bar p^\alpha \bar\nabla_\alpha \bar t_\nu}{\bar p \cdot \bar t} = \Omega^{-2} \left(\dot x^\mu - p^\mu - S^{\mu\nu} \nabla_\nu \log\Omega\right).
\end{equation}
Adding $\bar p^\mu$ cancels the last term and gives $\dd x^\mu/\dd\bar\lambda = \Omega^{-2} \dot x^\mu$. For the momentum equation, the conformal connection and $DS^{\mu\nu}/\dd\lambda = 2 p^{[\mu} \dot x^{\nu]}$ give the first equality below, while the conformal curvature transformation gives the second:
\begin{align}
    \Omega^2 \frac{\bar D \bar p_\mu}{\dd\bar\lambda} &- \frac{Dp_\mu}{\dd\lambda} =\nonumber\\
    &=S_\mu{}^\nu p^\alpha \Big(\nabla_\alpha \nabla_\nu \log\Omega  - \nabla_\alpha \log\Omega\,\nabla_\nu \log\Omega\Big) \nonumber\\
    &= -\frac{\Omega^2}{2} \bar R_{\mu\nu\alpha\beta} \bar p^\nu \bar S^{\alpha\beta} + \frac12 R_{\mu\nu\alpha\beta} p^\nu S^{\alpha\beta}.
\end{align}
The unbarred momentum equation therefore implies the barred one.

The conformal Killing--Yano tensor, its exterior derivative and divergence, and the associated conformal Killing tensor transform as
\begin{subequations}
\begin{align}
    \bar h &= \Omega^3 h, \\
    \dd\bar h &= 3 \Omega^2 \dd\Omega \wedge h,\\
    \frac13 \bar\nabla^\nu \bar h_{\nu\mu} &= \Omega \left(k_\mu + h_{\nu\mu} \nabla^\nu \log\Omega\right),\\
    \bar{\mathcal K}^{\mu\nu} &= \mathcal K^{\mu\nu}.
\end{align}
\end{subequations}
Since $\bar h$ is generally not closed, its hidden charge must use the full expression \eqref{eq:D_CKY}. The two contributions transform as
\begin{subequations}
\begin{align}
    \bar{\mathcal K}^{\mu\nu} \bar p_\mu \bar p_\nu &= \mathcal K^{\mu\nu} p_\mu p_\nu  - 2 S^{\alpha\beta} \mathcal K_{\beta\mu} p^\mu \nabla_\alpha \log\Omega,\\
    \bar{\mathcal L}_{\alpha\beta\mu} \bar S^{\alpha\beta} \bar p^\mu &= \mathcal L_{\alpha\beta\mu} S^{\alpha\beta} p^\mu  + 2 S^{\alpha\beta} \mathcal K_{\beta\mu} p^\mu \nabla_\alpha \log\Omega.
\end{align}
\end{subequations}
The additional terms cancel, giving $\bar{\mathcal D} = \mathcal D$. For $\kappa = k,m$, the momentum and spin contributions likewise cancel:
\begin{equation}
    \bar{\mathcal C}_\kappa - \mathcal C_\kappa = \kappa_\alpha \left(S^{\alpha\beta} + S^{\beta\alpha}\right) \nabla_\beta \log\Omega = 0.
\end{equation}
Both vectors remain Killing because $\Omega$ depends only on $r,y$, so $\bar E = E$ and $\bar L = L$.

To construct a conformally Carter solution, reconstruct the Carter solution on either branch and apply \cref{eq:conformal_ray_map}. Given barred initial data, first invert this map and then perform the appropriate Carter reduction, including the leading-principal extension in \cref{app:principal_observer} when the inverse map produces transverse momentum components of order $\varepsilon$. When the inverse map is evaluated as a truncated expression, we choose an exactly null representative of the resulting momentum, differing from it only at order $\varepsilon^2$, before computing the auxiliary geodesic constants. This is particularly important for leading-principal data, where the auxiliary hidden constant is itself of order $\varepsilon^2$. The separation parameter satisfies
\begin{equation}
    \frac{\dd\gamma}{\dd\bar\lambda} = \begin{cases}
    \displaystyle \frac{N}{\Omega^2}, & \text{nonprincipal branch},\\[6pt]
    \displaystyle \frac{1}{\Omega^2 \Sigma(\hat x)} \frac{\dd\hat\lambda}{\dd\lambda}, & \text{principal branch},
    \end{cases}
\end{equation}
where the principal clock is given by \cref{eq:app_principal_affine_clock}. The conformal factor is evaluated at the physical centroid. The spacetime curve is unchanged, but the momentum correction in \cref{eq:conformal_ray_map} must be included when computing its measured direction.

\section{Polarization-dependent black hole shadows}

In this section, we derive the momentum and velocity descriptions of the shadow boundary from the conserved quantities, and provide the explicit shadow boundary equations in Kerr. We retain terms through first order in $\varepsilon$ and use the outgoing radial branch at an observer outside the contributing spherical rays. Unless indicated otherwise, all quantities are evaluated at the physical centroid, and the observed directions are obtained at $(r_O,y_O)$.

\subsection{Momentum components and the shadow boundary equations}
\label{app:shadow_momentum}

We first solve the conservation laws for the physical momentum in terms of $E,L,\mathcal D$, and then derive the celestial-angle equations and endpoint condition in the main text. We use $P = r^2 E - L$, $Q = y^2 E + L$, and $\mathcal R = P^2 + \Delta_r \mathcal D$, with $E > 0$, $\mathcal D < 0$, and $P > 0$ at the observer. Primes denote derivatives of the corresponding metric function. Substituting the Carter observer and spin tensor into the Killing charges gives
\begin{subequations}

\begin{align}
    P &= \Delta_r(p^\tau + y^2 p^\psi) - \varepsilon s\,\frac{r p^y + y p^r}{p^\tau + y^2 p^\psi},\\
    Q &= -\Delta_y(p^\tau - r^2 p^\psi) \nonumber\\
    &\quad + \frac{\varepsilon s}{\Delta_r(p^\tau + y^2 p^\psi)} \left[r \Delta_r p^y + \left(y \Delta_y - \frac{\Sigma \Delta_y'}2\right)p^r\right].
\end{align}
\end{subequations}
The leading relation $p^\tau + y^2 p^\psi = P/\Delta_r$ may be used inside the spin corrections. The hidden constant, simplified using the null constraint and these Killing-charge relations, has the equivalent radial and angular forms
\begin{subequations}
\begin{align}
    \mathcal D &= -\frac{P^2}{\Delta_r} + \frac{\Sigma^2(p^r)^2}{\Delta_r} + \frac{2 \varepsilon s E y \Sigma}{P} p^r, \\
    &= -\frac{\Sigma^2(p^y)^2 + Q^2}{\Delta_y} + \frac{2 \varepsilon s r(\Delta_r Q + P \Delta_y)}{P \Delta_y} p^y \nonumber\\
    &\quad + \frac{\varepsilon s \Sigma p^r}{P} \left(4 E y - \frac{Q \Delta_y'}{\Delta_y}\right). 
\end{align}
\end{subequations}
Completing the radial square gives $(\Sigma p^r + \varepsilon s E y \Delta_r/P)^2 = \mathcal R$. On the outgoing radial branch relevant to the exterior observer,
\begin{equation}
    p^r = \frac{\sqrt{\mathcal R}}{\Sigma} - \frac{\varepsilon s E y \Delta_r}{P \Sigma}.
\end{equation}
Substituting the leading radial momentum into the angular correction and completing the polar square gives
\begin{align}
    &\left[\Sigma p^y - \frac{\varepsilon s r(\Delta_r Q + P \Delta_y)}{P \Sigma}\right]^2 = -\Delta_y \mathcal D - Q^2 \nonumber\\
    &\qquad + \frac{\varepsilon s \sqrt{\mathcal R}}{P}(4 E y \Delta_y - Q \Delta_y') + \mathcal O(\varepsilon^2).
    \label{eq:shadow_polar_quadratic}
\end{align}
Solving the Killing-charge relations for the remaining components and contracting with the Carter coframe gives
\begin{subequations}

\begin{align}
    \sqrt{\Sigma}\,p^0 &= \frac{P}{\sqrt{\Delta_r}} + \frac{\varepsilon s \sqrt{\Delta_r}}{P} \left(r p^y + \frac{y \sqrt{\mathcal R}}{\Sigma}\right),\\
    \sqrt{\Sigma}\,p^1 &= \frac{\sqrt{\mathcal R}}{\sqrt{\Delta_r}} - \frac{\varepsilon s E y \sqrt{\Delta_r}}{P},\\
    \sqrt{\Sigma}\,p^2 &= \frac{\Sigma p^y}{\sqrt{\Delta_y}},\\
    \sqrt{\Sigma}\,p^3 &= -\frac{Q}{\sqrt{\Delta_y}} + \frac{\varepsilon s}{P \sqrt{\Delta_y}} \Bigg[r \Delta_r p^y \nonumber\\
    &\qquad + \left(y \Delta_y - \frac{\Sigma \Delta_y'}2\right) \frac{\sqrt{\mathcal R}}{\Sigma}\Bigg].
\end{align}
\end{subequations}
Here $p^y$ satisfies \cref{eq:shadow_polar_quadratic}, and the coordinate momentum follows from $p = p^a e_a$.

For the celestial angles, the transverse momentum norm simplifies to
\begin{align}
    \Sigma \big[(p^2)^2 + (p^3)^2\big] &= -\mathcal D + 2 \varepsilon s \Bigg[r p^y + \frac{y \sqrt{\mathcal R}}{P \Sigma}(2 P + Q)\Bigg].
\end{align}
The leading relation
\begin{equation}
    \Sigma p^y = \sqrt{-\Delta_y \mathcal D} \cos\beta + \mathcal O(\varepsilon)
    \label{eq:app_shadow_leading_polar}
\end{equation}
may be used inside first-order terms. Substitution into $\sin\beta = -p^3/\sqrt{(p^2)^2 + (p^3)^2}$ gives
\begin{align}
    &\sin\beta = \frac{Q}{\sqrt{-\Delta_y \mathcal D}} + \frac{\varepsilon s r}{\Sigma} \left(\frac{Q}{\mathcal D} - \frac{\Delta_r}{P}\right) \cos\beta \nonumber\\
    &\quad + \frac{\varepsilon s \sqrt{\mathcal R}}{P \sqrt{-\Delta_y \mathcal D}} \Bigg[\frac{\Delta_y'}2 - \frac{y \Delta_y}{\Sigma}  + \frac{y Q(2 P + Q)}{\Sigma \mathcal D}\Bigg].
\end{align}
Moving the term proportional to $\cos\beta$ to the left and retaining the remaining normalization in square-root form gives \cref{eq:shadow_azimuth}. Likewise,
\begin{equation}
    \cos\alpha = \frac{\sqrt{\mathcal R} - \varepsilon s E y \Delta_r/P}{P + \dfrac{\varepsilon s \Delta_r}{P} \left(r p^y + \dfrac{y \sqrt{\mathcal R}}{\Sigma}\right)}.
\end{equation}
Expanding this ratio with $E \Sigma = P + Q$ and \cref{eq:app_shadow_leading_polar} gives \cref{eq:shadow_polar_angle}. Evaluating both results at $(r_O,y_O)$ produces the shadow equation in the main text.

The two angular branches of the shadow boundary meet where the right-hand side of \cref{eq:shadow_azimuth} has unit magnitude, giving the endpoint condition \cref{eq:shadow_endpoint}. Expanding this condition through first order and using $Q^2 = -\Delta_y \mathcal D$ at leading order gives their common polar momentum,
\begin{equation}
    \Sigma p^y = \frac{\varepsilon s r(\Delta_r Q + P \Delta_y)}{P \Sigma} + \mathcal O(\varepsilon^2).
\end{equation}
The quantities multiplying $\varepsilon$ may be evaluated at the leading-order endpoint. For numerical evaluation, the momentum at the observer is reconstructed from \cref{eq:shadow_momentum_decomposition} using the angular branches and the normalization \eqref{eq:shadow_observer_frequency}. We do not separately solve the truncated quadratic relation \eqref{eq:shadow_polar_quadratic} near its merging roots, where the omitted terms can affect the momentum at first order.

\subsection{Directions defined by the centroid velocity}
\label{app:shadow_velocity}

We derive the celestial-angle equations obtained from the centroid velocity $v^a = (e^a)_\mu \dot x^\mu$ and compare their directions and admissible spherical radii with the momentum construction. Define $(\alpha_v,\beta_v)$ by the convention in \cref{eq:shadow_momentum_decomposition}. The supplementary condition gives $v^0 = p^0$, while $p \cdot (v - p) = 0$ implies $v \cdot v = \mathcal O(\varepsilon^2)$. Thus $v^i/v^0$ is a unit spatial direction through first order. A common parameter factor cancels from all angular ratios, so we use $\dd x^\mu/\dd\gamma$ below.

Contracting the Carter-observer position equation with the coframe and substituting the conserved quantities gives the three components needed for these ratios:
\begin{subequations}

\begin{align}
    \sqrt{\frac{\Delta_r}{\Sigma}} (e^0)_\mu \frac{\dd x^\mu}{\dd\gamma} &= P + \frac{2 \varepsilon s y \Delta_r \sqrt{\mathcal R}}{P \Sigma},\\
    \sqrt{\frac{\Delta_r}{\Sigma}} (e^1)_\mu \frac{\dd x^\mu}{\dd\gamma} &= \sqrt{\mathcal R},\\
    -\sqrt{\frac{\Delta_y}{\Sigma}} (e^3)_\mu \frac{\dd x^\mu}{\dd\gamma} &= Q + \varepsilon s \Bigg[\Sigma \left(\frac{\Delta_r'}{2 P} - \frac{2 r E \Delta_r}{P^2}\right) p^y \nonumber\\
    &\quad + \frac{\sqrt{\mathcal R}}{P} \left(\frac{\Delta_y'}2 - \frac{2 y \Delta_y}{\Sigma}\right)\Bigg].
\end{align}
\end{subequations}
Nullness determines the transverse norm,
\begin{equation}
    \sum_{a=2}^3 \left[(e^a)_\mu \frac{\dd x^\mu}{\dd\gamma}\right]^2 = -\Sigma \mathcal D + 4 \varepsilon s y \sqrt{\mathcal R}.
\end{equation}
Its ratio to the squared time component gives $\sin^2\alpha_v$. In the azimuthal ratio, using $\Sigma p^y = \sqrt{-\Delta_y \mathcal D} \cos\beta_v + \mathcal O(\varepsilon)$ turns the term proportional to $p^y$ into a shift of $\beta_v$. Hence
\begin{subequations}
\label{eq:shadow_velocity_curve}
\begin{align}
    \sin\alpha_v &= \left.\frac{\sqrt{\Delta_r \left(-\mathcal D + \dfrac{4 \varepsilon s y \sqrt{\mathcal R}}{\Sigma}\right)}}{P + \dfrac{2 \varepsilon s y \Delta_r \sqrt{\mathcal R}}{P \Sigma}}\right|_{(r_O,y_O)},\\
    &\left.\sin\left[\beta_v - \varepsilon s \left(\frac{\Delta_r'}{2 P} - \frac{2 r E \Delta_r}{P^2}\right)\right]\right|_{(r_O,y_O)} \nonumber\\
    &\quad = \left.\frac{Q + \dfrac{\varepsilon s \sqrt{\mathcal R}}{P} \left(\dfrac{\Delta_y'}2 - \dfrac{2 y \Delta_y}{\Sigma}\right)}{\sqrt{\Delta_y \left(-\mathcal D + \dfrac{4 \varepsilon s y \sqrt{\mathcal R}}{\Sigma}\right)}}\right|_{(r_O,y_O)}.
\end{align}
\end{subequations}
The outgoing branch has $\alpha_v < \pi/2$. The critical ratios remain those in \cref{eq:shadow_critical_constants}. Requiring the last right-hand side to have magnitude at most one gives
\begin{equation}
    \left[Q^2 + \Delta_y \mathcal D - \frac{\varepsilon s \sqrt{\mathcal R}}{P}(4 E y \Delta_y - Q \Delta_y')\right]_{(r_O,y_O)} \leq 0.
\end{equation}
Here, $P + Q = E \Sigma$ was used, and the inequality was expanded through first order. Its equality agrees with \cref{eq:shadow_endpoint} through that order: the velocity and momentum constructions have the same first-order endpoint condition, but generally give different celestial curves at finite distance.

Since $v^0 = p^0$, the spatial components of the two sky directions differ by $-(v^i - p^i)/p^0$ through first order. If $\Delta_r = r^2 - 2 M r + \mathcal O(1)$ and $\Delta_r' = 2 r - 2 M + \mathcal O(r^{-1})$, a regular shadow family with bounded critical ratios has $v^i - p^i = \mathcal O(\varepsilon s r_O^{-2})$ at fixed angular position. With $p^0 = \varepsilon \omega_O$, the directional difference is therefore $\mathcal O(\omega_O^{-1} r_O^{-2})$.

\subsection{Kerr shadows}
\label{app:shadow_kerr}

We specialize the general shadow equations to Kerr, first in terms of the celestial angles and then in terms of the projected coordinates used in the plots. For $\Delta_r = r^2 - 2 M r + a^2$, the critical ratios in \cref{eq:shadow_critical_constants} become
\begin{subequations}

\begin{align}
    \frac{L}{E} &= r_{\rm p}^2 - \frac{2 r_{\rm p}(r_{\rm p}^2 - 2 M r_{\rm p} + a^2)}{r_{\rm p} - M}, \\
    \frac{\mathcal D}{E^2} &= -\frac{4 r_{\rm p}^2(r_{\rm p}^2 - 2 M r_{\rm p} + a^2)}{(r_{\rm p} - M)^2}.
\end{align}
\end{subequations}
We use $E = E^{(0)}$ from \cref{eq:shadow_leading_energy}. This leading energy is sufficient to impose the fixed observed frequency in the first-order angular corrections. Together with $P = r^2 E - L$, $Q = y^2 E + L$, and $\mathcal R = P^2 + \Delta_r \mathcal D$, these expressions determine the observer quantities as functions of $r_{\rm p}$. All quantities below are evaluated at $r = r_O$ and $y = y_O = a \cos\theta_O$, with $0 < \theta_O < \pi$, $\Delta_y = a^2 - y^2 > 0$, and $\Sigma = r^2 + y^2$.

To specify the curves used in the figures, we evaluate the first-order angular relations without further expansion of their squares, square roots, or trigonometric functions. The following equations define this numerical prescription; retaining these combinations does not determine the physical corrections beyond first order.

For each allowed spherical radius $r_{\rm p}$, the azimuthal relation has two solutions. Taking the principal branch of $\arcsin$, these solutions and the corresponding polar angles are
\begin{subequations}
\label{eq:shadow_kerr_angles}
\begin{align}
    \beta_\pm(r_{\rm p}) &= \frac{\pi}{2}(1 \mp 1) - \frac{\varepsilon s r}{\Sigma}\left(\frac{\Delta_r}{P} - \frac{Q}{\mathcal D}\right) \nonumber\\
    &\quad \pm \arcsin\!\left[\frac{Q - \dfrac{\varepsilon s y(r^2 + a^2)\sqrt{\mathcal R}}{P\Sigma}}{\sqrt{\Delta_y\left[-\mathcal D + \dfrac{2\varepsilon s y\sqrt{\mathcal R}}{P\Sigma}(2 P + Q)\right]}}\right], \label{eq:shadow_kerr_azimuth}
\end{align}
\begin{align}
    \cos\alpha_\pm(r_{\rm p}) &= \frac{\sqrt{\mathcal R}}{P} - \frac{\varepsilon s\Delta_r}{P^2\Sigma}\Bigg[y\left(P + Q + \frac{\mathcal R}{P}\right) \nonumber\\
    &\qquad + \frac{r\sqrt{-\Delta_y\mathcal D\mathcal R}}{P}\cos\beta_\pm(r_{\rm p})\Bigg]. 
\end{align}
\end{subequations}
The azimuths are understood modulo $2\pi$, and the upper and lower signs are paired consistently. These expressions solve \cref{eq:shadow_momentum_curve} after substituting the Kerr metric functions. In the geodesic limit $\varepsilon s = 0$, they reproduce the Kerr celestial-angle formulas in \cite[Eq.~(24)]{Grenzebach2014}.

The spherical radius parametrizes the two branches of the projected boundary. Applying \cref{eq:shadow_orthogonal_projection} without re-expanding gives the plotted coordinates
\begin{subequations}
\label{eq:shadow_kerr_curve}
\begin{align}
    X_\pm(r_{\rm p}) &= -\sqrt{1 - \cos^2\alpha_\pm(r_{\rm p})}\,\sin\beta_\pm(r_{\rm p}), \\
    Y_\pm(r_{\rm p}) &= -\sqrt{1 - \cos^2\alpha_\pm(r_{\rm p})}\,\cos\beta_\pm(r_{\rm p}).
\end{align}
\end{subequations}
Thus, the critical ratios, leading energy, and \cref{eq:shadow_kerr_angles,eq:shadow_kerr_curve} provide an explicit parametrization entirely in terms of $r_{\rm p}$.

The spherical rays must be unstable, outside the outer horizon, inside the observer, and accessible along the outgoing radial branch. In particular, $P > 0$ and $\mathcal R \geq 0$ at the observer, and the displayed angular and projected expressions must be real. Requiring the argument of $\arcsin$ in \cref{eq:shadow_kerr_azimuth} to lie in $[-1,1]$ gives
\begin{align}
    &\left[Q - \frac{\varepsilon s y(r^2 + a^2)\sqrt{\mathcal R}}{P\Sigma}\right]^2 \nonumber\\
    &\quad \leq \Delta_y\left[-\mathcal D + \frac{2\varepsilon s y\sqrt{\mathcal R}}{P\Sigma}(2 P + Q)\right],
\end{align}
with a positive right-hand side. Under the single-boundary assumption, the two relevant equality roots determine $r_{\rm p}^{\min}$ and $r_{\rm p}^{\max}$. This squared condition is evaluated as written, including the square of the spin correction in its left-hand side.

At either endpoint, the argument of $\arcsin$ is $+1$ or $-1$, so $\beta_+$ and $\beta_-$ coincide modulo $2\pi$. Consequently, $X_+ = X_-$ and $Y_+ = Y_-$ there. Following the upper-sign branch from $r_{\rm p}^{\min}$ to $r_{\rm p}^{\max}$ and the lower-sign branch in the reverse direction produces the closed boundary. The same construction applies to the nearly polar observer at $\theta_O = \pi/1000$; the narrow interval of spherical radii is resolved using increased numerical precision.

\bibliography{references}

%apsrev4-2.bst 2019-01-14 (MD) hand-edited version of apsrev4-1.bst
%Control: key (0)
%Control: author (8) initials jnrlst
%Control: editor formatted (1) identically to author
%Control: production of article title (0) allowed
%Control: page (0) single
%Control: year (1) truncated
%Control: production of eprint (0) enabled
\begin{thebibliography}{62}%
\makeatletter
\providecommand \@ifxundefined [1]{%
 \@ifx{#1\undefined}
}%
\providecommand \@ifnum [1]{%
 \ifnum #1\expandafter \@firstoftwo
 \else \expandafter \@secondoftwo
 \fi
}%
\providecommand \@ifx [1]{%
 \ifx #1\expandafter \@firstoftwo
 \else \expandafter \@secondoftwo
 \fi
}%
\providecommand \natexlab [1]{#1}%
\providecommand \enquote  [1]{``#1''}%
\providecommand \bibnamefont  [1]{#1}%
\providecommand \bibfnamefont [1]{#1}%
\providecommand \citenamefont [1]{#1}%
\providecommand \href@noop [0]{\@secondoftwo}%
\providecommand \href [0]{\begingroup \@sanitize@url \@href}%
\providecommand \@href[1]{\@@startlink{#1}\@@href}%
\providecommand \@@href[1]{\endgroup#1\@@endlink}%
\providecommand \@sanitize@url [0]{\catcode `\\12\catcode `\$12\catcode `\&12\catcode `\#12\catcode `\^12\catcode `\_12\catcode `\%12\relax}%
\providecommand \@@startlink[1]{}%
\providecommand \@@endlink[0]{}%
\providecommand \url  [0]{\begingroup\@sanitize@url \@url }%
\providecommand \@url [1]{\endgroup\@href {#1}{\urlprefix }}%
\providecommand \urlprefix  [0]{URL }%
\providecommand \Eprint [0]{\href }%
\providecommand \doibase [0]{https://doi.org/}%
\providecommand \selectlanguage [0]{\@gobble}%
\providecommand \bibinfo  [0]{\@secondoftwo}%
\providecommand \bibfield  [0]{\@secondoftwo}%
\providecommand \translation [1]{[#1]}%
\providecommand \BibitemOpen [0]{}%
\providecommand \bibitemStop [0]{}%
\providecommand \bibitemNoStop [0]{.\EOS\space}%
\providecommand \EOS [0]{\spacefactor3000\relax}%
\providecommand \BibitemShut  [1]{\csname bibitem#1\endcsname}%
\let\auto@bib@innerbib\@empty
%</preamble>
\bibitem [{\citenamefont {Sinova}\ \emph {et~al.}(2015)\citenamefont {Sinova}, \citenamefont {Valenzuela}, \citenamefont {Wunderlich}, \citenamefont {Back},\ and\ \citenamefont {Jungwirth}}]{SHE_review}%
  \BibitemOpen
  \bibfield  {author} {\bibinfo {author} {\bibfnamefont {J.}~\bibnamefont {Sinova}}, \bibinfo {author} {\bibfnamefont {S.~O.}\ \bibnamefont {Valenzuela}}, \bibinfo {author} {\bibfnamefont {J.}~\bibnamefont {Wunderlich}}, \bibinfo {author} {\bibfnamefont {C.~H.}\ \bibnamefont {Back}},\ and\ \bibinfo {author} {\bibfnamefont {T.}~\bibnamefont {Jungwirth}},\ }\bibfield  {title} {\bibinfo {title} {{Spin Hall effects}},\ }\href {https://doi.org/10.1103/RevModPhys.87.1213} {\bibfield  {journal} {\bibinfo  {journal} {Reviews of Modern Physics}\ }\textbf {\bibinfo {volume} {87}},\ \bibinfo {pages} {1213} (\bibinfo {year} {2015})}\BibitemShut {NoStop}%
\bibitem [{\citenamefont {Sinova}\ \emph {et~al.}(2004)\citenamefont {Sinova}, \citenamefont {Culcer}, \citenamefont {Niu}, \citenamefont {Sinitsyn}, \citenamefont {Jungwirth},\ and\ \citenamefont {MacDonald}}]{PhysRevLett.92.126603}%
  \BibitemOpen
  \bibfield  {author} {\bibinfo {author} {\bibfnamefont {J.}~\bibnamefont {Sinova}}, \bibinfo {author} {\bibfnamefont {D.}~\bibnamefont {Culcer}}, \bibinfo {author} {\bibfnamefont {Q.}~\bibnamefont {Niu}}, \bibinfo {author} {\bibfnamefont {N.~A.}\ \bibnamefont {Sinitsyn}}, \bibinfo {author} {\bibfnamefont {T.}~\bibnamefont {Jungwirth}},\ and\ \bibinfo {author} {\bibfnamefont {A.~H.}\ \bibnamefont {MacDonald}},\ }\bibfield  {title} {\bibinfo {title} {Universal intrinsic spin hall effect},\ }\href {https://doi.org/10.1103/PhysRevLett.92.126603} {\bibfield  {journal} {\bibinfo  {journal} {Physical Review Letters}\ }\textbf {\bibinfo {volume} {92}},\ \bibinfo {pages} {126603} (\bibinfo {year} {2004})}\BibitemShut {NoStop}%
\bibitem [{\citenamefont {Dyakonov}\ and\ \citenamefont {Perel}(1971{\natexlab{a}})}]{originalSHE1}%
  \BibitemOpen
  \bibfield  {author} {\bibinfo {author} {\bibfnamefont {M.~I.}\ \bibnamefont {Dyakonov}}\ and\ \bibinfo {author} {\bibfnamefont {V.~I.}\ \bibnamefont {Perel}},\ }\bibfield  {title} {\bibinfo {title} {{Possibility of orienting electron spins with current}},\ }\href@noop {} {\bibfield  {journal} {\bibinfo  {journal} {Soviet Journal of Experimental and Theoretical Physics Letters}\ }\textbf {\bibinfo {volume} {13}},\ \bibinfo {pages} {467} (\bibinfo {year} {1971}{\natexlab{a}})}\BibitemShut {NoStop}%
\bibitem [{\citenamefont {Dyakonov}\ and\ \citenamefont {Perel}(1971{\natexlab{b}})}]{originalSHE2}%
  \BibitemOpen
  \bibfield  {author} {\bibinfo {author} {\bibfnamefont {M.~I.}\ \bibnamefont {Dyakonov}}\ and\ \bibinfo {author} {\bibfnamefont {V.~I.}\ \bibnamefont {Perel}},\ }\bibfield  {title} {\bibinfo {title} {{Current-induced spin orientation of electrons in semiconductors}},\ }\href {https://doi.org/https://doi.org/10.1016/0375-9601(71)90196-4} {\bibfield  {journal} {\bibinfo  {journal} {Physics Letters A}\ }\textbf {\bibinfo {volume} {35}},\ \bibinfo {pages} {459} (\bibinfo {year} {1971}{\natexlab{b}})}\BibitemShut {NoStop}%
\bibitem [{\citenamefont {{Bakun}}\ \emph {et~al.}(1984)\citenamefont {{Bakun}}, \citenamefont {{Zakharchenya}}, \citenamefont {{Rogachev}}, \citenamefont {{Tkachuk}},\ and\ \citenamefont {{Fle{\v \i}sher}}}]{originalSHE3}%
  \BibitemOpen
  \bibfield  {author} {\bibinfo {author} {\bibfnamefont {A.~A.}\ \bibnamefont {{Bakun}}}, \bibinfo {author} {\bibfnamefont {B.~P.}\ \bibnamefont {{Zakharchenya}}}, \bibinfo {author} {\bibfnamefont {A.~A.}\ \bibnamefont {{Rogachev}}}, \bibinfo {author} {\bibfnamefont {M.~N.}\ \bibnamefont {{Tkachuk}}},\ and\ \bibinfo {author} {\bibfnamefont {V.~G.}\ \bibnamefont {{Fle{\v \i}sher}}},\ }\bibfield  {title} {\bibinfo {title} {{Observation of a surface photocurrent caused by optical orientation of electrons in a semiconductor}},\ }\href@noop {} {\bibfield  {journal} {\bibinfo  {journal} {Soviet Journal of Experimental and Theoretical Physics Letters}\ }\textbf {\bibinfo {volume} {40}},\ \bibinfo {pages} {1293} (\bibinfo {year} {1984})}\BibitemShut {NoStop}%
\bibitem [{\citenamefont {Kato}\ \emph {et~al.}(2004)\citenamefont {Kato}, \citenamefont {Myers}, \citenamefont {Gossard},\ and\ \citenamefont {Awschalom}}]{originalSHE4}%
  \BibitemOpen
  \bibfield  {author} {\bibinfo {author} {\bibfnamefont {Y.~K.}\ \bibnamefont {Kato}}, \bibinfo {author} {\bibfnamefont {R.~C.}\ \bibnamefont {Myers}}, \bibinfo {author} {\bibfnamefont {A.~C.}\ \bibnamefont {Gossard}},\ and\ \bibinfo {author} {\bibfnamefont {D.~D.}\ \bibnamefont {Awschalom}},\ }\bibfield  {title} {\bibinfo {title} {{Observation of the spin Hall effect in semiconductors}},\ }\href {https://doi.org/10.1126/science.1105514} {\bibfield  {journal} {\bibinfo  {journal} {Science}\ }\textbf {\bibinfo {volume} {306}},\ \bibinfo {pages} {1910} (\bibinfo {year} {2004})}\BibitemShut {NoStop}%
\bibitem [{\citenamefont {Bliokh}\ \emph {et~al.}(2015)\citenamefont {Bliokh}, \citenamefont {Rodr\'{i}guez-Fortu\~{n}o}, \citenamefont {Nori},\ and\ \citenamefont {Zayats}}]{SOI_review}%
  \BibitemOpen
  \bibfield  {author} {\bibinfo {author} {\bibfnamefont {K.~Y.}\ \bibnamefont {Bliokh}}, \bibinfo {author} {\bibfnamefont {F.~J.}\ \bibnamefont {Rodr\'{i}guez-Fortu\~{n}o}}, \bibinfo {author} {\bibfnamefont {F.}~\bibnamefont {Nori}},\ and\ \bibinfo {author} {\bibfnamefont {A.~V.}\ \bibnamefont {Zayats}},\ }\bibfield  {title} {\bibinfo {title} {{Spin-orbit interactions of light}},\ }\href {https://doi.org/10.1038/nphoton.2015.201} {\bibfield  {journal} {\bibinfo  {journal} {Nature Photonics}\ }\textbf {\bibinfo {volume} {9}},\ \bibinfo {pages} {796} (\bibinfo {year} {2015})}\BibitemShut {NoStop}%
\bibitem [{\citenamefont {Ling}\ \emph {et~al.}(2017)\citenamefont {Ling}, \citenamefont {Zhou}, \citenamefont {Huang}, \citenamefont {Liu}, \citenamefont {Qiu}, \citenamefont {Luo},\ and\ \citenamefont {Wen}}]{SHEL_review}%
  \BibitemOpen
  \bibfield  {author} {\bibinfo {author} {\bibfnamefont {X.}~\bibnamefont {Ling}}, \bibinfo {author} {\bibfnamefont {X.}~\bibnamefont {Zhou}}, \bibinfo {author} {\bibfnamefont {K.}~\bibnamefont {Huang}}, \bibinfo {author} {\bibfnamefont {Y.}~\bibnamefont {Liu}}, \bibinfo {author} {\bibfnamefont {C.-W.}\ \bibnamefont {Qiu}}, \bibinfo {author} {\bibfnamefont {H.}~\bibnamefont {Luo}},\ and\ \bibinfo {author} {\bibfnamefont {S.}~\bibnamefont {Wen}},\ }\bibfield  {title} {\bibinfo {title} {{Recent advances in the spin Hall effect of light}},\ }\href {http://stacks.iop.org/0034-4885/80/i=6/a=066401} {\bibfield  {journal} {\bibinfo  {journal} {Reports on Progress in Physics}\ }\textbf {\bibinfo {volume} {80}},\ \bibinfo {pages} {066401} (\bibinfo {year} {2017})}\BibitemShut {NoStop}%
\bibitem [{\citenamefont {Liberman}\ and\ \citenamefont {Zel'dovich}(1992)}]{SHE-L_original}%
  \BibitemOpen
  \bibfield  {author} {\bibinfo {author} {\bibfnamefont {V.~S.}\ \bibnamefont {Liberman}}\ and\ \bibinfo {author} {\bibfnamefont {B.~Y.}\ \bibnamefont {Zel'dovich}},\ }\bibfield  {title} {\bibinfo {title} {{Spin-orbit interaction of a photon in an inhomogeneous medium}},\ }\href {https://doi.org/10.1103/PhysRevA.46.5199} {\bibfield  {journal} {\bibinfo  {journal} {Physical Review A}\ }\textbf {\bibinfo {volume} {46}},\ \bibinfo {pages} {5199} (\bibinfo {year} {1992})}\BibitemShut {NoStop}%
\bibitem [{\citenamefont {Onoda}\ \emph {et~al.}(2004)\citenamefont {Onoda}, \citenamefont {Murakami},\ and\ \citenamefont {Nagaosa}}]{SHE_original}%
  \BibitemOpen
  \bibfield  {author} {\bibinfo {author} {\bibfnamefont {M.}~\bibnamefont {Onoda}}, \bibinfo {author} {\bibfnamefont {S.}~\bibnamefont {Murakami}},\ and\ \bibinfo {author} {\bibfnamefont {N.}~\bibnamefont {Nagaosa}},\ }\bibfield  {title} {\bibinfo {title} {Hall effect of light},\ }\href {https://doi.org/10.1103/PhysRevLett.93.083901} {\bibfield  {journal} {\bibinfo  {journal} {Physical Review Letters}\ }\textbf {\bibinfo {volume} {93}},\ \bibinfo {pages} {083901} (\bibinfo {year} {2004})}\BibitemShut {NoStop}%
\bibitem [{\citenamefont {Bliokh}(2009)}]{Bliokh2009}%
  \BibitemOpen
  \bibfield  {author} {\bibinfo {author} {\bibfnamefont {K.~Y.}\ \bibnamefont {Bliokh}},\ }\bibfield  {title} {\bibinfo {title} {{Geometrodynamics of polarized light: Berry phase and spin Hall effect in a gradient-index medium}},\ }\href {https://doi.org/10.1088/1464-4258/11/9/094009} {\bibfield  {journal} {\bibinfo  {journal} {Journal of Optics A: Pure and Applied Optics}\ }\textbf {\bibinfo {volume} {11}},\ \bibinfo {pages} {094009} (\bibinfo {year} {2009})}\BibitemShut {NoStop}%
\bibitem [{\citenamefont {Duval}\ \emph {et~al.}(2006)\citenamefont {Duval}, \citenamefont {Horv\'ath},\ and\ \citenamefont {Horv\'athy}}]{Duval2006}%
  \BibitemOpen
  \bibfield  {author} {\bibinfo {author} {\bibfnamefont {C.}~\bibnamefont {Duval}}, \bibinfo {author} {\bibfnamefont {Z.}~\bibnamefont {Horv\'ath}},\ and\ \bibinfo {author} {\bibfnamefont {P.~A.}\ \bibnamefont {Horv\'athy}},\ }\bibfield  {title} {\bibinfo {title} {Fermat principle for spinning light},\ }\href {https://doi.org/10.1103/PhysRevD.74.021701} {\bibfield  {journal} {\bibinfo  {journal} {Physical Review D}\ }\textbf {\bibinfo {volume} {74}},\ \bibinfo {pages} {021701} (\bibinfo {year} {2006})}\BibitemShut {NoStop}%
\bibitem [{\citenamefont {Duval}\ \emph {et~al.}(2007)\citenamefont {Duval}, \citenamefont {Horv\'ath},\ and\ \citenamefont {Horv\'athy}}]{Duval2007}%
  \BibitemOpen
  \bibfield  {author} {\bibinfo {author} {\bibfnamefont {C.}~\bibnamefont {Duval}}, \bibinfo {author} {\bibfnamefont {Z.}~\bibnamefont {Horv\'ath}},\ and\ \bibinfo {author} {\bibfnamefont {P.~A.}\ \bibnamefont {Horv\'athy}},\ }\bibfield  {title} {\bibinfo {title} {{Geometrical spinoptics and the optical Hall effect}},\ }\href {https://doi.org/https://doi.org/10.1016/j.geomphys.2006.07.003} {\bibfield  {journal} {\bibinfo  {journal} {Journal of Geometry and Physics}\ }\textbf {\bibinfo {volume} {57}},\ \bibinfo {pages} {925} (\bibinfo {year} {2007})}\BibitemShut {NoStop}%
\bibitem [{\citenamefont {Bliokh}\ \emph {et~al.}(2008)\citenamefont {Bliokh}, \citenamefont {Niv}, \citenamefont {Kleiner},\ and\ \citenamefont {Hasman}}]{Bliokh2008}%
  \BibitemOpen
  \bibfield  {author} {\bibinfo {author} {\bibfnamefont {K.~Y.}\ \bibnamefont {Bliokh}}, \bibinfo {author} {\bibfnamefont {A.}~\bibnamefont {Niv}}, \bibinfo {author} {\bibfnamefont {V.}~\bibnamefont {Kleiner}},\ and\ \bibinfo {author} {\bibfnamefont {E.}~\bibnamefont {Hasman}},\ }\bibfield  {title} {\bibinfo {title} {{Geometrodynamics of spinning light}},\ }\href {http://dx.doi.org/10.1038/nphoton.2008.229} {\bibfield  {journal} {\bibinfo  {journal} {Nature Photonics}\ }\textbf {\bibinfo {volume} {2}},\ \bibinfo {pages} {748} (\bibinfo {year} {2008})}\BibitemShut {NoStop}%
\bibitem [{\citenamefont {Hosten}\ and\ \citenamefont {Kwiat}(2008)}]{SHEL_experiment}%
  \BibitemOpen
  \bibfield  {author} {\bibinfo {author} {\bibfnamefont {O.}~\bibnamefont {Hosten}}\ and\ \bibinfo {author} {\bibfnamefont {P.}~\bibnamefont {Kwiat}},\ }\bibfield  {title} {\bibinfo {title} {{Observation of the spin Hall effect of light via weak measurements}},\ }\href {https://doi.org/10.1126/science.1152697} {\bibfield  {journal} {\bibinfo  {journal} {Science}\ }\textbf {\bibinfo {volume} {319}},\ \bibinfo {pages} {787} (\bibinfo {year} {2008})}\BibitemShut {NoStop}%
\bibitem [{\citenamefont {Dooghin}\ \emph {et~al.}(1992)\citenamefont {Dooghin}, \citenamefont {Kundikova}, \citenamefont {Liberman},\ and\ \citenamefont {Zel'dovich}}]{OpticalMagnus}%
  \BibitemOpen
  \bibfield  {author} {\bibinfo {author} {\bibfnamefont {A.~V.}\ \bibnamefont {Dooghin}}, \bibinfo {author} {\bibfnamefont {N.~D.}\ \bibnamefont {Kundikova}}, \bibinfo {author} {\bibfnamefont {V.~S.}\ \bibnamefont {Liberman}},\ and\ \bibinfo {author} {\bibfnamefont {B.~Y.}\ \bibnamefont {Zel'dovich}},\ }\bibfield  {title} {\bibinfo {title} {{Optical Magnus effect}},\ }\href {https://doi.org/10.1103/PhysRevA.45.8204} {\bibfield  {journal} {\bibinfo  {journal} {Physical Review A}\ }\textbf {\bibinfo {volume} {45}},\ \bibinfo {pages} {8204} (\bibinfo {year} {1992})}\BibitemShut {NoStop}%
\bibitem [{\citenamefont {Collingbourne}\ \emph {et~al.}(2026)\citenamefont {Collingbourne}, \citenamefont {Oancea},\ and\ \citenamefont {Sbierski}}]{SHE_math}%
  \BibitemOpen
  \bibfield  {author} {\bibinfo {author} {\bibfnamefont {S.~C.}\ \bibnamefont {Collingbourne}}, \bibinfo {author} {\bibfnamefont {M.~A.}\ \bibnamefont {Oancea}},\ and\ \bibinfo {author} {\bibfnamefont {J.}~\bibnamefont {Sbierski}},\ }\bibfield  {title} {\bibinfo {title} {{A mathematical description of the spin Hall effect of light in inhomogeneous media}},\ }\href {https://arxiv.org/abs/2603.28697} {\bibfield  {journal} {\bibinfo  {journal} {arXiv:2603.28697}\ } (\bibinfo {year} {2026})}\BibitemShut {NoStop}%
\bibitem [{\citenamefont {Qin}\ \emph {et~al.}(2009)\citenamefont {Qin}, \citenamefont {Li}, \citenamefont {He},\ and\ \citenamefont {Gong}}]{Qin:09}%
  \BibitemOpen
  \bibfield  {author} {\bibinfo {author} {\bibfnamefont {Y.}~\bibnamefont {Qin}}, \bibinfo {author} {\bibfnamefont {Y.}~\bibnamefont {Li}}, \bibinfo {author} {\bibfnamefont {H.}~\bibnamefont {He}},\ and\ \bibinfo {author} {\bibfnamefont {Q.}~\bibnamefont {Gong}},\ }\bibfield  {title} {\bibinfo {title} {Measurement of spin hall effect of reflected light},\ }\href {https://doi.org/10.1364/OL.34.002551} {\bibfield  {journal} {\bibinfo  {journal} {Optics Letters}\ }\textbf {\bibinfo {volume} {34}},\ \bibinfo {pages} {2551} (\bibinfo {year} {2009})}\BibitemShut {NoStop}%
\bibitem [{\citenamefont {Chakravarthy}\ and\ \citenamefont {Viswanathan}(2019)}]{PradeepChakravarthy:19}%
  \BibitemOpen
  \bibfield  {author} {\bibinfo {author} {\bibfnamefont {T.~P.}\ \bibnamefont {Chakravarthy}}\ and\ \bibinfo {author} {\bibfnamefont {N.~K.}\ \bibnamefont {Viswanathan}},\ }\bibfield  {title} {\bibinfo {title} {Direct and reciprocal spin-orbit interaction effects in a graded-index medium},\ }\href {https://doi.org/10.1364/OSAC.2.001576} {\bibfield  {journal} {\bibinfo  {journal} {OSA Continuum}\ }\textbf {\bibinfo {volume} {2}},\ \bibinfo {pages} {1576} (\bibinfo {year} {2019})}\BibitemShut {NoStop}%
\bibitem [{\citenamefont {L{\"o}ffler}\ \emph {et~al.}(2010)\citenamefont {L{\"o}ffler}, \citenamefont {{van Exter}}, \citenamefont {{'t Hooft}}, \citenamefont {Eliel}, \citenamefont {Hermans}, \citenamefont {Broer},\ and\ \citenamefont {Woerdman}}]{LOFFLER20103367}%
  \BibitemOpen
  \bibfield  {author} {\bibinfo {author} {\bibfnamefont {W.}~\bibnamefont {L{\"o}ffler}}, \bibinfo {author} {\bibfnamefont {M.}~\bibnamefont {{van Exter}}}, \bibinfo {author} {\bibfnamefont {G.}~\bibnamefont {{'t Hooft}}}, \bibinfo {author} {\bibfnamefont {E.}~\bibnamefont {Eliel}}, \bibinfo {author} {\bibfnamefont {K.}~\bibnamefont {Hermans}}, \bibinfo {author} {\bibfnamefont {D.}~\bibnamefont {Broer}},\ and\ \bibinfo {author} {\bibfnamefont {J.}~\bibnamefont {Woerdman}},\ }\bibfield  {title} {\bibinfo {title} {Polarization-dependent goos--h{\"a}nchen shift at a graded dielectric interface},\ }\href {https://doi.org/https://doi.org/10.1016/j.optcom.2010.04.055} {\bibfield  {journal} {\bibinfo  {journal} {Optics Communications}\ }\textbf {\bibinfo {volume} {283}},\ \bibinfo {pages} {3367} (\bibinfo {year} {2010})}\BibitemShut {NoStop}%
\bibitem [{\citenamefont {Andersson}\ and\ \citenamefont {Oancea}(2023)}]{GSHE_rev}%
  \BibitemOpen
  \bibfield  {author} {\bibinfo {author} {\bibfnamefont {L.}~\bibnamefont {Andersson}}\ and\ \bibinfo {author} {\bibfnamefont {M.~A.}\ \bibnamefont {Oancea}},\ }\bibfield  {title} {\bibinfo {title} {{Spin Hall effects in the sky}},\ }\href {https://doi.org/10.1088/1361-6382/ace021} {\bibfield  {journal} {\bibinfo  {journal} {Classical and Quantum Gravity}\ }\textbf {\bibinfo {volume} {40}},\ \bibinfo {pages} {154002} (\bibinfo {year} {2023})}\BibitemShut {NoStop}%
\bibitem [{\citenamefont {Oancea}\ \emph {et~al.}(2020)\citenamefont {Oancea}, \citenamefont {Joudioux}, \citenamefont {Dodin}, \citenamefont {Ruiz}, \citenamefont {Paganini},\ and\ \citenamefont {Andersson}}]{GSHE2020}%
  \BibitemOpen
  \bibfield  {author} {\bibinfo {author} {\bibfnamefont {M.~A.}\ \bibnamefont {Oancea}}, \bibinfo {author} {\bibfnamefont {J.}~\bibnamefont {Joudioux}}, \bibinfo {author} {\bibfnamefont {I.~Y.}\ \bibnamefont {Dodin}}, \bibinfo {author} {\bibfnamefont {D.~E.}\ \bibnamefont {Ruiz}}, \bibinfo {author} {\bibfnamefont {C.~F.}\ \bibnamefont {Paganini}},\ and\ \bibinfo {author} {\bibfnamefont {L.}~\bibnamefont {Andersson}},\ }\bibfield  {title} {\bibinfo {title} {{Gravitational spin Hall effect of light}},\ }\href {https://doi.org/10.1103/PhysRevD.102.024075} {\bibfield  {journal} {\bibinfo  {journal} {Physical Review D}\ }\textbf {\bibinfo {volume} {102}},\ \bibinfo {pages} {024075} (\bibinfo {year} {2020})}\BibitemShut {NoStop}%
\bibitem [{\citenamefont {Harte}\ and\ \citenamefont {Oancea}(2022)}]{HarteOancea}%
  \BibitemOpen
  \bibfield  {author} {\bibinfo {author} {\bibfnamefont {A.~I.}\ \bibnamefont {Harte}}\ and\ \bibinfo {author} {\bibfnamefont {M.~A.}\ \bibnamefont {Oancea}},\ }\bibfield  {title} {\bibinfo {title} {Spin {H}all effects and the localization of massless spinning particles},\ }\href {https://doi.org/10.1103/PhysRevD.105.104061} {\bibfield  {journal} {\bibinfo  {journal} {Physical Review D}\ }\textbf {\bibinfo {volume} {105}},\ \bibinfo {pages} {104061} (\bibinfo {year} {2022})}\BibitemShut {NoStop}%
\bibitem [{\citenamefont {Gosselin}\ \emph {et~al.}(2007{\natexlab{a}})\citenamefont {Gosselin}, \citenamefont {B{\'e}rard},\ and\ \citenamefont {Mohrbach}}]{SHE_QM1}%
  \BibitemOpen
  \bibfield  {author} {\bibinfo {author} {\bibfnamefont {P.}~\bibnamefont {Gosselin}}, \bibinfo {author} {\bibfnamefont {A.}~\bibnamefont {B{\'e}rard}},\ and\ \bibinfo {author} {\bibfnamefont {H.}~\bibnamefont {Mohrbach}},\ }\bibfield  {title} {\bibinfo {title} {{Spin Hall effect of photons in a static gravitational field}},\ }\href {https://doi.org/10.1103/PhysRevD.75.084035} {\bibfield  {journal} {\bibinfo  {journal} {Physical Review D}\ }\textbf {\bibinfo {volume} {75}},\ \bibinfo {pages} {084035} (\bibinfo {year} {2007}{\natexlab{a}})}\BibitemShut {NoStop}%
\bibitem [{\citenamefont {Frolov}(2020)}]{Frolov2020}%
  \BibitemOpen
  \bibfield  {author} {\bibinfo {author} {\bibfnamefont {V.~P.}\ \bibnamefont {Frolov}},\ }\bibfield  {title} {\bibinfo {title} {Maxwell equations in a curved spacetime: Spin optics approximation},\ }\href {https://doi.org/10.1103/PhysRevD.102.084013} {\bibfield  {journal} {\bibinfo  {journal} {Physical Review D}\ }\textbf {\bibinfo {volume} {102}},\ \bibinfo {pages} {084013} (\bibinfo {year} {2020})}\BibitemShut {NoStop}%
\bibitem [{\citenamefont {Frolov}(2024{\natexlab{a}})}]{PhysRevD.110.064020}%
  \BibitemOpen
  \bibfield  {author} {\bibinfo {author} {\bibfnamefont {V.~P.}\ \bibnamefont {Frolov}},\ }\bibfield  {title} {\bibinfo {title} {Spinoptics in a curved spacetime},\ }\href {https://doi.org/10.1103/PhysRevD.110.064020} {\bibfield  {journal} {\bibinfo  {journal} {Physical Review D}\ }\textbf {\bibinfo {volume} {110}},\ \bibinfo {eid} {064020} (\bibinfo {year} {2024}{\natexlab{a}})}\BibitemShut {NoStop}%
\bibitem [{\citenamefont {Frolov}(2024{\natexlab{b}})}]{Frolov_2024a}%
  \BibitemOpen
  \bibfield  {author} {\bibinfo {author} {\bibfnamefont {V.~P.}\ \bibnamefont {Frolov}},\ }\bibfield  {title} {\bibinfo {title} {{Spinoptics in the Schwarzschild spacetime}},\ }\href {https://doi.org/10.1088/1475-7516/2024/12/051} {\bibfield  {journal} {\bibinfo  {journal} {Journal of Cosmology and Astroparticle Physics}\ }\textbf {\bibinfo {volume} {2024}}\bibinfo  {number} { (12)},\ \bibinfo {pages} {051}}\BibitemShut {NoStop}%
\bibitem [{\citenamefont {Andersson}\ \emph {et~al.}(2021)\citenamefont {Andersson}, \citenamefont {Joudioux}, \citenamefont {Oancea},\ and\ \citenamefont {Raj}}]{GSHE_GW}%
  \BibitemOpen
\bibfield  {number} {  }\bibfield  {author} {\bibinfo {author} {\bibfnamefont {L.}~\bibnamefont {Andersson}}, \bibinfo {author} {\bibfnamefont {J.}~\bibnamefont {Joudioux}}, \bibinfo {author} {\bibfnamefont {M.~A.}\ \bibnamefont {Oancea}},\ and\ \bibinfo {author} {\bibfnamefont {A.}~\bibnamefont {Raj}},\ }\bibfield  {title} {\bibinfo {title} {Propagation of polarized gravitational waves},\ }\href {https://doi.org/10.1103/PhysRevD.103.044053} {\bibfield  {journal} {\bibinfo  {journal} {Physical Review D}\ }\textbf {\bibinfo {volume} {103}},\ \bibinfo {pages} {044053} (\bibinfo {year} {2021})}\BibitemShut {NoStop}%
\bibitem [{\citenamefont {Yamamoto}(2018)}]{SHE_GW}%
  \BibitemOpen
  \bibfield  {author} {\bibinfo {author} {\bibfnamefont {N.}~\bibnamefont {Yamamoto}},\ }\bibfield  {title} {\bibinfo {title} {{Spin Hall effect of gravitational waves}},\ }\href {https://doi.org/10.1103/PhysRevD.98.061701} {\bibfield  {journal} {\bibinfo  {journal} {Physical Review D}\ }\textbf {\bibinfo {volume} {98}},\ \bibinfo {pages} {061701(R)} (\bibinfo {year} {2018})}\BibitemShut {NoStop}%
\bibitem [{\citenamefont {Frolov}\ and\ \citenamefont {Shoom}(2024)}]{Frolov_2024}%
  \BibitemOpen
  \bibfield  {author} {\bibinfo {author} {\bibfnamefont {V.~P.}\ \bibnamefont {Frolov}}\ and\ \bibinfo {author} {\bibfnamefont {A.~A.}\ \bibnamefont {Shoom}},\ }\bibfield  {title} {\bibinfo {title} {Gravitational spinoptics in a curved space-time},\ }\href {https://doi.org/10.1088/1475-7516/2024/10/039} {\bibfield  {journal} {\bibinfo  {journal} {Journal of Cosmology and Astroparticle Physics}\ }\textbf {\bibinfo {volume} {2024}}\bibinfo  {number} { (10)},\ \bibinfo {eid} {039}}\BibitemShut {NoStop}%
\bibitem [{\citenamefont {Audretsch}(1981)}]{audretsch}%
  \BibitemOpen
\bibfield  {number} {  }\bibfield  {author} {\bibinfo {author} {\bibfnamefont {J.}~\bibnamefont {Audretsch}},\ }\bibfield  {title} {\bibinfo {title} {{Trajectories and spin motion of massive spin-$\frac{1}{2}$ particles in gravitational fields}},\ }\href {https://doi.org/https://doi.org/10.1088/0305-4470/14/2/017} {\bibfield  {journal} {\bibinfo  {journal} {Journal of Physics A: Mathematical and General}\ }\textbf {\bibinfo {volume} {14}},\ \bibinfo {pages} {411} (\bibinfo {year} {1981})}\BibitemShut {NoStop}%
\bibitem [{\citenamefont {R{\"u}diger}(1981)}]{rudiger}%
  \BibitemOpen
  \bibfield  {author} {\bibinfo {author} {\bibfnamefont {R.}~\bibnamefont {R{\"u}diger}},\ }\bibfield  {title} {\bibinfo {title} {{The Dirac equation and spinning particles in general relativity}},\ }\href {https://doi.org/https://doi.org/10.1098/rspa.1981.0132} {\bibfield  {journal} {\bibinfo  {journal} {Proceedings of the Royal Society A: Mathematical, Physical and Engineering Sciences}\ }\textbf {\bibinfo {volume} {377}},\ \bibinfo {pages} {417} (\bibinfo {year} {1981})}\BibitemShut {NoStop}%
\bibitem [{\citenamefont {Oancea}\ and\ \citenamefont {Kumar}(2023)}]{GSHE_Dirac}%
  \BibitemOpen
  \bibfield  {author} {\bibinfo {author} {\bibfnamefont {M.~A.}\ \bibnamefont {Oancea}}\ and\ \bibinfo {author} {\bibfnamefont {A.}~\bibnamefont {Kumar}},\ }\bibfield  {title} {\bibinfo {title} {{Semiclassical analysis of Dirac fields on curved spacetime}},\ }\href {https://doi.org/10.1103/PhysRevD.107.044029} {\bibfield  {journal} {\bibinfo  {journal} {Physical Review D}\ }\textbf {\bibinfo {volume} {107}},\ \bibinfo {pages} {044029} (\bibinfo {year} {2023})}\BibitemShut {NoStop}%
\bibitem [{\citenamefont {Gosselin}\ \emph {et~al.}(2007{\natexlab{b}})\citenamefont {Gosselin}, \citenamefont {B{\'e}rard},\ and\ \citenamefont {Mohrbach}}]{SHE_Dirac}%
  \BibitemOpen
  \bibfield  {author} {\bibinfo {author} {\bibfnamefont {P.}~\bibnamefont {Gosselin}}, \bibinfo {author} {\bibfnamefont {A.}~\bibnamefont {B{\'e}rard}},\ and\ \bibinfo {author} {\bibfnamefont {H.}~\bibnamefont {Mohrbach}},\ }\bibfield  {title} {\bibinfo {title} {{Semiclassical dynamics of Dirac particles interacting with a static gravitational field}},\ }\href {https://doi.org/https://doi.org/10.1016/j.physleta.2007.04.022} {\bibfield  {journal} {\bibinfo  {journal} {Physics Letters A}\ }\textbf {\bibinfo {volume} {368}},\ \bibinfo {pages} {356} (\bibinfo {year} {2007}{\natexlab{b}})}\BibitemShut {NoStop}%
\bibitem [{\citenamefont {Oancea}\ \emph {et~al.}(2024{\natexlab{a}})\citenamefont {Oancea}, \citenamefont {Stiskalek},\ and\ \citenamefont {Zumalac\'arregui}}]{GSHE_lensing}%
  \BibitemOpen
  \bibfield  {author} {\bibinfo {author} {\bibfnamefont {M.~A.}\ \bibnamefont {Oancea}}, \bibinfo {author} {\bibfnamefont {R.}~\bibnamefont {Stiskalek}},\ and\ \bibinfo {author} {\bibfnamefont {M.}~\bibnamefont {Zumalac\'arregui}},\ }\bibfield  {title} {\bibinfo {title} {Frequency- and polarization-dependent lensing of gravitational waves in strong gravitational fields},\ }\href {https://doi.org/10.1103/PhysRevD.109.124045} {\bibfield  {journal} {\bibinfo  {journal} {Physical Review D}\ }\textbf {\bibinfo {volume} {109}},\ \bibinfo {pages} {124045} (\bibinfo {year} {2024}{\natexlab{a}})}\BibitemShut {NoStop}%
\bibitem [{\citenamefont {Oancea}\ \emph {et~al.}(2024{\natexlab{b}})\citenamefont {Oancea}, \citenamefont {Stiskalek},\ and\ \citenamefont {Zumalac\'arregui}}]{GSHE_lensing2}%
  \BibitemOpen
  \bibfield  {author} {\bibinfo {author} {\bibfnamefont {M.~A.}\ \bibnamefont {Oancea}}, \bibinfo {author} {\bibfnamefont {R.}~\bibnamefont {Stiskalek}},\ and\ \bibinfo {author} {\bibfnamefont {M.}~\bibnamefont {Zumalac\'arregui}},\ }\bibfield  {title} {\bibinfo {title} {Probing general relativistic spin–orbit coupling with gravitational waves from hierarchical triple systems},\ }\href {https://doi.org/10.1093/mnrasl/slae084} {\bibfield  {journal} {\bibinfo  {journal} {Monthly Notices of the Royal Astronomical Society: Letters}\ }\textbf {\bibinfo {volume} {535}},\ \bibinfo {pages} {L1} (\bibinfo {year} {2024}{\natexlab{b}})}\BibitemShut {NoStop}%
\bibitem [{\citenamefont {Kubota}\ \emph {et~al.}(2024)\citenamefont {Kubota}, \citenamefont {Arai},\ and\ \citenamefont {Mukohyama}}]{PhysRevD.109.044027}%
  \BibitemOpen
  \bibfield  {author} {\bibinfo {author} {\bibfnamefont {K.-i.}\ \bibnamefont {Kubota}}, \bibinfo {author} {\bibfnamefont {S.}~\bibnamefont {Arai}},\ and\ \bibinfo {author} {\bibfnamefont {S.}~\bibnamefont {Mukohyama}},\ }\bibfield  {title} {\bibinfo {title} {Spin optics for gravitational waves lensed by a rotating object},\ }\href {https://doi.org/10.1103/PhysRevD.109.044027} {\bibfield  {journal} {\bibinfo  {journal} {Physical Review D}\ }\textbf {\bibinfo {volume} {109}},\ \bibinfo {pages} {044027} (\bibinfo {year} {2024})}\BibitemShut {NoStop}%
\bibitem [{\citenamefont {Nazar{\'e}}\ \emph {et~al.}(2026)\citenamefont {Nazar{\'e}}, \citenamefont {Santos}, \citenamefont {Cardoso},\ and\ \citenamefont {Nat{\'a}rio}}]{Nazare2026}%
  \BibitemOpen
  \bibfield  {author} {\bibinfo {author} {\bibfnamefont {F.}~\bibnamefont {Nazar{\'e}}}, \bibinfo {author} {\bibfnamefont {J.~S.}\ \bibnamefont {Santos}}, \bibinfo {author} {\bibfnamefont {V.}~\bibnamefont {Cardoso}},\ and\ \bibinfo {author} {\bibfnamefont {J.}~\bibnamefont {Nat{\'a}rio}},\ }\bibfield  {title} {\bibinfo {title} {{Wave optical imaging of an oscillating electric dipole orbiting a black hole}},\ }\href {https://arxiv.org/abs/2609.10689} {\bibfield  {journal} {\bibinfo  {journal} {arXiv:2609.10689}\ } (\bibinfo {year} {2026})}\BibitemShut {NoStop}%
\bibitem [{\citenamefont {Andersson}\ \emph {et~al.}(2026)\citenamefont {Andersson}, \citenamefont {Gray},\ and\ \citenamefont {Oancea}}]{GSHE_int}%
  \BibitemOpen
  \bibfield  {author} {\bibinfo {author} {\bibfnamefont {L.}~\bibnamefont {Andersson}}, \bibinfo {author} {\bibfnamefont {F.}~\bibnamefont {Gray}},\ and\ \bibinfo {author} {\bibfnamefont {M.~A.}\ \bibnamefont {Oancea}},\ }\bibfield  {title} {\bibinfo {title} {Conserved quantities and integrability for massless spinning particles in general relativity},\ }\href {https://doi.org/10.1103/85dv-n8tr} {\bibfield  {journal} {\bibinfo  {journal} {Physical Review D}\ }\textbf {\bibinfo {volume} {113}},\ \bibinfo {pages} {064061} (\bibinfo {year} {2026})}\BibitemShut {NoStop}%
\bibitem [{\citenamefont {Skoup{\'y}}\ and\ \citenamefont {Witzany}(2025)}]{SkoupyWitzany2025}%
  \BibitemOpen
  \bibfield  {author} {\bibinfo {author} {\bibfnamefont {V.}~\bibnamefont {Skoup{\'y}}}\ and\ \bibinfo {author} {\bibfnamefont {V.}~\bibnamefont {Witzany}},\ }\bibfield  {title} {\bibinfo {title} {Analytic solution for the motion of spinning particles in kerr spacetime},\ }\href {https://doi.org/10.1103/PhysRevLett.134.171401} {\bibfield  {journal} {\bibinfo  {journal} {Physical Review Letters}\ }\textbf {\bibinfo {volume} {134}},\ \bibinfo {pages} {171401} (\bibinfo {year} {2025})}\BibitemShut {NoStop}%
\bibitem [{\citenamefont {Grenzebach}\ \emph {et~al.}(2014)\citenamefont {Grenzebach}, \citenamefont {Perlick},\ and\ \citenamefont {L\"ammerzahl}}]{Grenzebach2014}%
  \BibitemOpen
  \bibfield  {author} {\bibinfo {author} {\bibfnamefont {A.}~\bibnamefont {Grenzebach}}, \bibinfo {author} {\bibfnamefont {V.}~\bibnamefont {Perlick}},\ and\ \bibinfo {author} {\bibfnamefont {C.}~\bibnamefont {L\"ammerzahl}},\ }\bibfield  {title} {\bibinfo {title} {Photon regions and shadows of {Kerr}-{Newman}-{NUT} black holes with a cosmological constant},\ }\href {https://doi.org/10.1103/PhysRevD.89.124004} {\bibfield  {journal} {\bibinfo  {journal} {Physical Review D}\ }\textbf {\bibinfo {volume} {89}},\ \bibinfo {pages} {124004} (\bibinfo {year} {2014})}\BibitemShut {NoStop}%
\bibitem [{\citenamefont {Grenzebach}\ \emph {et~al.}(2015)\citenamefont {Grenzebach}, \citenamefont {Perlick},\ and\ \citenamefont {L\"ammerzahl}}]{BHS3}%
  \BibitemOpen
  \bibfield  {author} {\bibinfo {author} {\bibfnamefont {A.}~\bibnamefont {Grenzebach}}, \bibinfo {author} {\bibfnamefont {V.}~\bibnamefont {Perlick}},\ and\ \bibinfo {author} {\bibfnamefont {C.}~\bibnamefont {L\"ammerzahl}},\ }\bibfield  {title} {\bibinfo {title} {Photon regions and shadows of accelerated black holes},\ }\href {https://doi.org/10.1142/S0218271815420249} {\bibfield  {journal} {\bibinfo  {journal} {International Journal of Modern Physics D}\ }\textbf {\bibinfo {volume} {24}},\ \bibinfo {pages} {1542024} (\bibinfo {year} {2015})}\BibitemShut {NoStop}%
\bibitem [{\citenamefont {Perlick}\ and\ \citenamefont {Tsupko}(2022)}]{Perlick2021}%
  \BibitemOpen
  \bibfield  {author} {\bibinfo {author} {\bibfnamefont {V.}~\bibnamefont {Perlick}}\ and\ \bibinfo {author} {\bibfnamefont {O.~Y.}\ \bibnamefont {Tsupko}},\ }\bibfield  {title} {\bibinfo {title} {{Calculating black hole shadows: Review of analytical studies}},\ }\href {https://doi.org/https://doi.org/10.1016/j.physrep.2021.10.004} {\bibfield  {journal} {\bibinfo  {journal} {Physics Reports}\ }\textbf {\bibinfo {volume} {947}},\ \bibinfo {pages} {1} (\bibinfo {year} {2022})}\BibitemShut {NoStop}%
\bibitem [{\citenamefont {{Dixon}}(1974)}]{Dixon74}%
  \BibitemOpen
  \bibfield  {author} {\bibinfo {author} {\bibfnamefont {W.~G.}\ \bibnamefont {{Dixon}}},\ }\bibfield  {title} {\bibinfo {title} {{Dynamics of extended bodies in general relativity. III. Equations of motion}},\ }\href {https://doi.org/10.1098/rsta.1974.0046} {\bibfield  {journal} {\bibinfo  {journal} {Philosophical Transactions of the Royal Society A: Mathematical, Physical and Engineering Sciences}\ }\textbf {\bibinfo {volume} {277}},\ \bibinfo {pages} {59} (\bibinfo {year} {1974})}\BibitemShut {NoStop}%
\bibitem [{\citenamefont {Harte}(2012)}]{HarteGrav}%
  \BibitemOpen
  \bibfield  {author} {\bibinfo {author} {\bibfnamefont {A.~I.}\ \bibnamefont {Harte}},\ }\bibfield  {title} {\bibinfo {title} {Mechanics of extended masses in general relativity},\ }\href {https://doi.org/10.1088/0264-9381/29/5/055012} {\bibfield  {journal} {\bibinfo  {journal} {Classical and Quantum Gravity}\ }\textbf {\bibinfo {volume} {29}},\ \bibinfo {pages} {055012} (\bibinfo {year} {2012})}\BibitemShut {NoStop}%
\bibitem [{\citenamefont {Frolov}\ \emph {et~al.}(2017)\citenamefont {Frolov}, \citenamefont {Krtou{\v s}},\ and\ \citenamefont {Kubiz{\v n}{\'a}k}}]{Frolov2017}%
  \BibitemOpen
  \bibfield  {author} {\bibinfo {author} {\bibfnamefont {V.~P.}\ \bibnamefont {Frolov}}, \bibinfo {author} {\bibfnamefont {P.}~\bibnamefont {Krtou{\v s}}},\ and\ \bibinfo {author} {\bibfnamefont {D.}~\bibnamefont {Kubiz{\v n}{\'a}k}},\ }\bibfield  {title} {\bibinfo {title} {Black holes, hidden symmetries, and complete integrability},\ }\href {https://doi.org/10.1007/s41114-017-0009-9} {\bibfield  {journal} {\bibinfo  {journal} {Living Reviews in Relativity}\ }\textbf {\bibinfo {volume} {20}},\ \bibinfo {pages} {6} (\bibinfo {year} {2017})}\BibitemShut {NoStop}%
\bibitem [{\citenamefont {Vines}\ \emph {et~al.}(2016)\citenamefont {Vines}, \citenamefont {Kunst}, \citenamefont {Steinhoff},\ and\ \citenamefont {Hinderer}}]{Vines2016}%
  \BibitemOpen
  \bibfield  {author} {\bibinfo {author} {\bibfnamefont {J.}~\bibnamefont {Vines}}, \bibinfo {author} {\bibfnamefont {D.}~\bibnamefont {Kunst}}, \bibinfo {author} {\bibfnamefont {J.}~\bibnamefont {Steinhoff}},\ and\ \bibinfo {author} {\bibfnamefont {T.}~\bibnamefont {Hinderer}},\ }\bibfield  {title} {\bibinfo {title} {{Canonical Hamiltonian for an extended test body in curved spacetime: To quadratic order in spin}},\ }\href {https://doi.org/https://doi.org/10.1103/PhysRevD.93.103008} {\bibfield  {journal} {\bibinfo  {journal} {Physical Review D}\ }\textbf {\bibinfo {volume} {93}},\ \bibinfo {pages} {103008} (\bibinfo {year} {2016})}\BibitemShut {NoStop}%
\bibitem [{\citenamefont {Poisson}\ \emph {et~al.}(2011)\citenamefont {Poisson}, \citenamefont {Pound},\ and\ \citenamefont {Vega}}]{Poisson2011}%
  \BibitemOpen
  \bibfield  {author} {\bibinfo {author} {\bibfnamefont {E.}~\bibnamefont {Poisson}}, \bibinfo {author} {\bibfnamefont {A.}~\bibnamefont {Pound}},\ and\ \bibinfo {author} {\bibfnamefont {I.}~\bibnamefont {Vega}},\ }\bibfield  {title} {\bibinfo {title} {The motion of point particles in curved spacetime},\ }\href {https://doi.org/10.12942/lrr-2011-7} {\bibfield  {journal} {\bibinfo  {journal} {Living Reviews in Relativity}\ }\textbf {\bibinfo {volume} {14}},\ \bibinfo {pages} {7} (\bibinfo {year} {2011})}\BibitemShut {NoStop}%
\bibitem [{\citenamefont {Hackmann}\ and\ \citenamefont {Lämmerzahl}(2014)}]{10.1063/1.4861945}%
  \BibitemOpen
  \bibfield  {author} {\bibinfo {author} {\bibfnamefont {E.}~\bibnamefont {Hackmann}}\ and\ \bibinfo {author} {\bibfnamefont {C.}~\bibnamefont {Lämmerzahl}},\ }\bibfield  {title} {\bibinfo {title} {Analytical solution methods for geodesic motion},\ }\href {https://doi.org/10.1063/1.4861945} {\bibfield  {journal} {\bibinfo  {journal} {AIP Conference Proceedings}\ }\textbf {\bibinfo {volume} {1577}},\ \bibinfo {pages} {78} (\bibinfo {year} {2014})}\BibitemShut {NoStop}%
\bibitem [{\citenamefont {Gralla}\ and\ \citenamefont {Lupsasca}(2020)}]{GrallaLupsasca2020}%
  \BibitemOpen
  \bibfield  {author} {\bibinfo {author} {\bibfnamefont {S.~E.}\ \bibnamefont {Gralla}}\ and\ \bibinfo {author} {\bibfnamefont {A.}~\bibnamefont {Lupsasca}},\ }\bibfield  {title} {\bibinfo {title} {{Null geodesics of the Kerr exterior}},\ }\href {https://doi.org/10.1103/PhysRevD.101.044032} {\bibfield  {journal} {\bibinfo  {journal} {Physical Review D}\ }\textbf {\bibinfo {volume} {101}},\ \bibinfo {pages} {044032} (\bibinfo {year} {2020})}\BibitemShut {NoStop}%
\bibitem [{\citenamefont {Hackmann}\ \emph {et~al.}(2010)\citenamefont {Hackmann}, \citenamefont {L{\"a}mmerzahl}, \citenamefont {Kagramanova},\ and\ \citenamefont {Kunz}}]{Hackmann2010}%
  \BibitemOpen
  \bibfield  {author} {\bibinfo {author} {\bibfnamefont {E.}~\bibnamefont {Hackmann}}, \bibinfo {author} {\bibfnamefont {C.}~\bibnamefont {L{\"a}mmerzahl}}, \bibinfo {author} {\bibfnamefont {V.}~\bibnamefont {Kagramanova}},\ and\ \bibinfo {author} {\bibfnamefont {J.}~\bibnamefont {Kunz}},\ }\bibfield  {title} {\bibinfo {title} {Analytical solution of the geodesic equation in kerr-(anti-) de sitter space-times},\ }\href {https://doi.org/10.1103/PhysRevD.81.044020} {\bibfield  {journal} {\bibinfo  {journal} {Physical Review D}\ }\textbf {\bibinfo {volume} {81}},\ \bibinfo {pages} {044020} (\bibinfo {year} {2010})}\BibitemShut {NoStop}%
\bibitem [{\citenamefont {Hackmann}\ \emph {et~al.}(2009)\citenamefont {Hackmann}, \citenamefont {Kagramanova}, \citenamefont {Kunz},\ and\ \citenamefont {L{\"a}mmerzahl}}]{Hackmann2009}%
  \BibitemOpen
  \bibfield  {author} {\bibinfo {author} {\bibfnamefont {E.}~\bibnamefont {Hackmann}}, \bibinfo {author} {\bibfnamefont {V.}~\bibnamefont {Kagramanova}}, \bibinfo {author} {\bibfnamefont {J.}~\bibnamefont {Kunz}},\ and\ \bibinfo {author} {\bibfnamefont {C.}~\bibnamefont {L{\"a}mmerzahl}},\ }\bibfield  {title} {\bibinfo {title} {Analytic solutions of the geodesic equation in axially symmetric space-times},\ }\href {https://doi.org/10.1209/0295-5075/88/30008} {\bibfield  {journal} {\bibinfo  {journal} {Europhysics Letters}\ }\textbf {\bibinfo {volume} {88}},\ \bibinfo {pages} {30008} (\bibinfo {year} {2009})}\BibitemShut {NoStop}%
\bibitem [{\citenamefont {Enolski}\ \emph {et~al.}(2011)\citenamefont {Enolski}, \citenamefont {Hackmann}, \citenamefont {Kagramanova}, \citenamefont {Kunz},\ and\ \citenamefont {L{\"a}mmerzahl}}]{ENOLSKI2011899}%
  \BibitemOpen
  \bibfield  {author} {\bibinfo {author} {\bibfnamefont {V.}~\bibnamefont {Enolski}}, \bibinfo {author} {\bibfnamefont {E.}~\bibnamefont {Hackmann}}, \bibinfo {author} {\bibfnamefont {V.}~\bibnamefont {Kagramanova}}, \bibinfo {author} {\bibfnamefont {J.}~\bibnamefont {Kunz}},\ and\ \bibinfo {author} {\bibfnamefont {C.}~\bibnamefont {L{\"a}mmerzahl}},\ }\bibfield  {title} {\bibinfo {title} {Inversion of hyperelliptic integrals of arbitrary genus with application to particle motion in general relativity},\ }\href {https://doi.org/https://doi.org/10.1016/j.geomphys.2011.01.001} {\bibfield  {journal} {\bibinfo  {journal} {Journal of Geometry and Physics}\ }\textbf {\bibinfo {volume} {61}},\ \bibinfo {pages} {899} (\bibinfo {year} {2011})}\BibitemShut {NoStop}%
\bibitem [{\citenamefont {Kubiz{\v n}{\'a}k}\ \emph {et~al.}(2009)\citenamefont {Kubiz{\v n}{\'a}k}, \citenamefont {Frolov}, \citenamefont {Krtou{\v s}},\ and\ \citenamefont {Connell}}]{Kubiznak2009}%
  \BibitemOpen
  \bibfield  {author} {\bibinfo {author} {\bibfnamefont {D.}~\bibnamefont {Kubiz{\v n}{\'a}k}}, \bibinfo {author} {\bibfnamefont {V.~P.}\ \bibnamefont {Frolov}}, \bibinfo {author} {\bibfnamefont {P.}~\bibnamefont {Krtou{\v s}}},\ and\ \bibinfo {author} {\bibfnamefont {P.}~\bibnamefont {Connell}},\ }\bibfield  {title} {\bibinfo {title} {Parallel-propagated frame along null geodesics in higher-dimensional black hole spacetimes},\ }\href {https://doi.org/10.1103/PhysRevD.79.024018} {\bibfield  {journal} {\bibinfo  {journal} {Physical Review D}\ }\textbf {\bibinfo {volume} {79}},\ \bibinfo {pages} {024018} (\bibinfo {year} {2009})}\BibitemShut {NoStop}%
\bibitem [{\citenamefont {Marck}(1983)}]{Marck1983Null}%
  \BibitemOpen
  \bibfield  {author} {\bibinfo {author} {\bibfnamefont {J.-A.}\ \bibnamefont {Marck}},\ }\bibfield  {title} {\bibinfo {title} {Parallel-tetrad on null geodesics in {Kerr-Newman} space-time},\ }\href {https://doi.org/10.1016/0375-9601(83)90197-4} {\bibfield  {journal} {\bibinfo  {journal} {Physics Letters A}\ }\textbf {\bibinfo {volume} {97}},\ \bibinfo {pages} {140} (\bibinfo {year} {1983})}\BibitemShut {NoStop}%
\bibitem [{\citenamefont {Cunha}\ and\ \citenamefont {Herdeiro}(2018)}]{BHS7}%
  \BibitemOpen
  \bibfield  {author} {\bibinfo {author} {\bibfnamefont {P.~V.~P.}\ \bibnamefont {Cunha}}\ and\ \bibinfo {author} {\bibfnamefont {C.~A.~R.}\ \bibnamefont {Herdeiro}},\ }\bibfield  {title} {\bibinfo {title} {{Shadows and strong gravitational lensing: A brief review}},\ }\href {https://doi.org/10.1007/s10714-018-2361-9} {\bibfield  {journal} {\bibinfo  {journal} {General Relativity and Gravitation}\ }\textbf {\bibinfo {volume} {50}},\ \bibinfo {pages} {42} (\bibinfo {year} {2018})}\BibitemShut {NoStop}%
\bibitem [{\citenamefont {Frolov}\ and\ \citenamefont {Koek}(2025)}]{Frolov2025}%
  \BibitemOpen
  \bibfield  {author} {\bibinfo {author} {\bibfnamefont {V.~P.}\ \bibnamefont {Frolov}}\ and\ \bibinfo {author} {\bibfnamefont {A.}~\bibnamefont {Koek}},\ }\bibfield  {title} {\bibinfo {title} {Spinoptics in the kerr spacetime: Polarized wave scattering},\ }\href {https://doi.org/10.1103/trh5-4sgq} {\bibfield  {journal} {\bibinfo  {journal} {Physical Review D}\ }\textbf {\bibinfo {volume} {111}},\ \bibinfo {pages} {104081} (\bibinfo {year} {2025})}\BibitemShut {NoStop}%
\bibitem [{\citenamefont {{The Event Horizon Telescope Collaboration}}(2019)}]{EHT2019}%
  \BibitemOpen
  \bibfield  {author} {\bibinfo {author} {\bibnamefont {{The Event Horizon Telescope Collaboration}}},\ }\bibfield  {title} {\bibinfo {title} {{First M87 Event Horizon Telescope Results. I. The Shadow of the Supermassive Black Hole}},\ }\href {https://doi.org/10.3847/2041-8213/ab0ec7} {\bibfield  {journal} {\bibinfo  {journal} {The Astrophysical Journal Letters}\ }\textbf {\bibinfo {volume} {875}},\ \bibinfo {pages} {L1} (\bibinfo {year} {2019})}\BibitemShut {NoStop}%
\bibitem [{\citenamefont {{The Event Horizon Telescope Collaboration}}(2022)}]{EHT2022}%
  \BibitemOpen
  \bibfield  {author} {\bibinfo {author} {\bibnamefont {{The Event Horizon Telescope Collaboration}}},\ }\bibfield  {title} {\bibinfo {title} {{First Sagittarius A* Event Horizon Telescope Results. I. The Shadow of the Supermassive Black Hole in the Center of the Milky Way}},\ }\href {https://doi.org/10.3847/2041-8213/ac6674} {\bibfield  {journal} {\bibinfo  {journal} {The Astrophysical Journal Letters}\ }\textbf {\bibinfo {volume} {930}},\ \bibinfo {pages} {L12} (\bibinfo {year} {2022})}\BibitemShut {NoStop}%
\bibitem [{\citenamefont {Dolan}(2008)}]{Dolan2008}%
  \BibitemOpen
  \bibfield  {author} {\bibinfo {author} {\bibfnamefont {S.~R.}\ \bibnamefont {Dolan}},\ }\bibfield  {title} {\bibinfo {title} {Scattering and absorption of gravitational plane waves by rotating black holes},\ }\href {https://doi.org/10.1088/0264-9381/25/23/235002} {\bibfield  {journal} {\bibinfo  {journal} {Classical and Quantum Gravity}\ }\textbf {\bibinfo {volume} {25}},\ \bibinfo {pages} {235002} (\bibinfo {year} {2008})}\BibitemShut {NoStop}%
\bibitem [{\citenamefont {Leite}\ \emph {et~al.}(2017)\citenamefont {Leite}, \citenamefont {Dolan},\ and\ \citenamefont {Crispino}}]{leite2017absorption}%
  \BibitemOpen
  \bibfield  {author} {\bibinfo {author} {\bibfnamefont {L.~C.~S.}\ \bibnamefont {Leite}}, \bibinfo {author} {\bibfnamefont {S.~R.}\ \bibnamefont {Dolan}},\ and\ \bibinfo {author} {\bibfnamefont {L.~C.~B.}\ \bibnamefont {Crispino}},\ }\bibfield  {title} {\bibinfo {title} {{Absorption of electromagnetic and gravitational waves by Kerr black holes}},\ }\href {https://doi.org/10.1016/j.physletb.2017.09.048} {\bibfield  {journal} {\bibinfo  {journal} {Physics Letters B}\ }\textbf {\bibinfo {volume} {774}},\ \bibinfo {pages} {130} (\bibinfo {year} {2017})}\BibitemShut {NoStop}%
\bibitem [{\citenamefont {Leite}\ \emph {et~al.}(2018)\citenamefont {Leite}, \citenamefont {Dolan},\ and\ \citenamefont {Crispino}}]{leite2018absorption}%
  \BibitemOpen
  \bibfield  {author} {\bibinfo {author} {\bibfnamefont {L.~C.~S.}\ \bibnamefont {Leite}}, \bibinfo {author} {\bibfnamefont {S.~R.}\ \bibnamefont {Dolan}},\ and\ \bibinfo {author} {\bibfnamefont {L.~C.~B.}\ \bibnamefont {Crispino}},\ }\bibfield  {title} {\bibinfo {title} {Absorption of electromagnetic plane waves by rotating black holes},\ }\href {https://doi.org/10.1103/PhysRevD.98.024046} {\bibfield  {journal} {\bibinfo  {journal} {Physical Review D}\ }\textbf {\bibinfo {volume} {98}},\ \bibinfo {pages} {024046} (\bibinfo {year} {2018})}\BibitemShut {NoStop}%
\end{thebibliography}%
\end{document}